\documentclass
[aps,amssymb,amsmath,floats,showpacs,pre,superscriptaddress]{revtex4}%
\usepackage{amsfonts}
\usepackage{amsmath}
\usepackage{amssymb}
\usepackage{epsf}
\usepackage{graphicx}
\usepackage{graphics}%
\begin{document}
\title{Laplacian spectra of complex networks and random walks on them: \\Are scale-free architectures really important?}
\author{A. N. Samukhin}
\email{samukhin@fis.ua.pt}
\affiliation{Departamento de F{\'{\i}}sica da Universidade de Aveiro, 3810-193 Aveiro, Portugal}
\affiliation{A. F. Ioffe Physico-Technical Institute, 194021 St. Petersburg, Russia}
\author{S. N. Dorogovtsev}
\email{sdorogov@fis.ua.pt}
\affiliation{Departamento de F{\'{\i}}sica da Universidade de Aveiro, 3810-193 Aveiro, Portugal}
\affiliation{A. F. Ioffe Physico-Technical Institute, 194021 St. Petersburg, Russia}
\author{J. F. F. Mendes}
\email{jfmendes@fis.ua.pt}
\affiliation{Departamento de F{\'{\i}}sica da Universidade de Aveiro, 3810-193 Aveiro, Portugal}

\begin{abstract}
We study the Laplacian operator of an uncorrelated random network and, as an
application, consider hopping processes (diffusion, random walks, signal
propagation, etc.) on 
networks. We develop a strict approach to these
problems. We derive an exact closed set of integral equations, which provide
the averages of the Laplacian operator's resolvent. This enables us to
describe the propagation of a signal and random walks on the network. We show
that the determining parameter in this problem is the minimum degree $q_{m}$
of vertices in the network and that the high-degree part of the degree
distribution is not that essential. The position of the lower edge of the Laplacian
spectrum $\lambda_{c}$ appears to be the same as in the regular Bethe lattice
with the coordination number $q_{m}$. Namely, $\lambda_{c}>0$ if $q_{m}>2$,
and $\lambda_{c}=0$ if $q_{m}\leq2$. In both these cases the density of
eigenvalues $\rho\left(  \lambda\right)  \rightarrow0$ as $\lambda
\rightarrow\lambda_{c}+0$, but the limiting behaviors near $\lambda_{c}$ are
very different. In terms of a distance from a starting vertex, the hopping
propagator is a steady moving Gaussian, broadening with time. This picture
qualitatively coincides with that for a regular Bethe lattice. Our analytical
results include the spectral density $\rho(\lambda)$ near $\lambda_{c}$ and
the long-time asymptotics of the autocorrelator and the propagator.

\end{abstract}
\maketitle

\section{Introduction}

\label{introduction}

The Laplacian spectra of random networks determine a wide circle of processes
taking place on these networks, see, e.g.,
\cite{sg07,ktr07,br88,jm,clv03,m91,kk07,aj04,m07} and references therein. Random
walks, signal propagation, synchronization, and many others are among these
processes. This is why the problem of Laplacian spectra of random networks
(especially, of its low-eigenvalue part which determines the long-time
behavior of relevant processes) is considered as one of central problems of
graph theory and the science of complex networks. In this paper we essentially
resolve this problem applying the strict statistical mechanics approach to
uncorrelated random networks with arbitrary degree distributions. These random
graphs constitute a basic class of complex networks.

One should note that leading contributions to the spectra and the asymptotics
of the random walk autocorrelator were found by Bray and Rodgers in 1988 in
the particular case of the Erd\H{o}s-R\'enyi graphs \cite{br88}. 
It is important that these classical graphs necessarily have dead ends and vertices with two connections. We will show that the absence of these vertices in a network qualitatively changes the spectra and the random walk asymptotics.   
Random walks
on hierarchically organized, deterministic, scale-free graphs were studied by
Noh and Rieger in Ref.~\cite{nr04}. Due to a very specific organization of
these graphs, their results are not applicable to equilibrium networks. This
is also the case in respect of the recent numerical work of Kujawski,
Tadi\'{c} and Rodgers \cite{ktr07}, who found the autocorrelator of a random
walk on a growing scale-free network by performing extensive numerical
simulations. Their network was strongly correlated in contrast to the
configuration model of a random graph, which we use in this work.  

For the sake of clearness, let us remind basic notions and terms for random networks. For more
detail see \cite{ab01a,dm01c,dmbook02,n03,blmch,dgm07,bb98}. A graph is completely
defined by its $N\times N$ \emph{adjacency matrix }$\hat{A}$, whose elements
$A_{ij}$ are the numbers of edges between $i$ and $j$. The vertex degree of
vertex $i$ is the number of edges, attached to this vertex: $q_{i}=\sum
_{j=1}^{N}A_{ij}=\sum_{j=1}^{N}A_{ji}$. In random networks, $q_{i}$ is a
random variable with \emph{a degree distribution} $\Pi\left(  q\right)
=\left\langle \delta\left(  q-q_{i}\right)  \right\rangle $.

In traditional mathematical models, $\Pi(q)$ is a rapidly decaying function
with a well-defined scale. For example, in the Erd\H{o}s-R\'{e}nyi model
\cite{er59}, 
which is a standard one, $\Pi(q)$ is
a Poisson distribution  
decaying as $\left(  \bar{q}/eq\right)  ^{q}$,
i.e., faster than any exponent. In contrast to these models, in most of
real-world networks degree distributions are heavy tailed. After the 
work \cite{ajb00d}, they are usually approximated by a power-law $\sim
q^{-\gamma}$ in the range of sufficiently high degrees. 
Note 
that 
the validity of this fitting is limited because real-world networks are
small (even the WWW has only about $10^{10}$ vertices), and so high degrees
are not observable. It is commonly believed that the \textquotedblleft
scale-free networks\textquotedblright\ are greatly distinguished from the
others in every aspect. This widespread belief actually implies a division of
all networks into two classes: \textquotedblleft scale-free
networks\textquotedblright\ and all others. In contrast to these beliefs, we
here show that 
scale-free (or, more generally,
heavy tailed) architectures of networks are not essential 
for a lower edge of the Laplacian spectra and the long-time
behavior of random walks characteristics. 
The resulting
dependences are determined by the minimum degree of vertices in a network. 
Heavy tails determine some coefficients and amplitudes but not a type of these singularities.

\begin{table*}[t] 
\centering
\caption{Asymptotics of the Laplacian spectral density $\rho\left(  \lambda\right)  $,
autocorrelator $\bar{P}_{0}\left(  t\right)  $ and propagator $\bar{P}_{l}\left(  t\right)  $ for the random uncorrelated
networks where $\Pi\left(q_m\right)$ is essentially distinct from $0$ and $1$.
Here $p_{l}^{\left(  eq\right)  }=P_{l}\left(  t\rightarrow\infty\right)  $\ are
stationary values of the correlator given by Eq.~(\protect\ref{235}) for $l=0$ and Eq. (\protect\ref{300}) otherwise.
$\beta=\pi\left( q_m-1 \right)^{1/4}\ln\left( q_m-1 \right)$.
The values of the parameters
in the pre-exponential factors are $\xi=9/10$ and $4/3$, $\eta=-7/30$ and $1/18$, and
$\zeta=13/30$ and $5/18$ for $q_m=1$ and $2$,
respectively. $v$ and $D$ are determined by the full form of the degree
distribution $\Pi\left( q \right)$. 
}%
\begin{tabular}
[c]{l|c|c}\hline
&  & \\[-10pt]
& Minimum vertex degree $q_{m}>2$ & Minimum vertex degree $q_{m}=1$ or
$2$\\[-10pt]
&  & \\\hline
&  & \\[-6pt]%
Spectral edge $\lambda_{c}$ & $\mathrm{\ }q_{m}-2\sqrt{q_{m}-1}$ & $0$\\[-6pt]
&  & \\\hline
&  & \\[-6pt]%
Spectral density & $\phantom{o}\exp\left[  \frac{\beta}{2\sqrt{\lambda
-\lambda_{c}}}-d\exp\left(  \frac{\beta}{\sqrt{\lambda-\lambda_{c}}}\right)
\right]  $, Eq. (\ref{390})$\phantom{o}$ & $p_{0}^{\left(  \mathrm{eq}\right)
}\delta\left(  \lambda\right)  +\mathrm{const~}\lambda^{-\xi}\exp\left(
-a/\sqrt{\lambda}\right)  $, Eqs. (\ref{470}), (\ref{540})\\[-6pt]
&  & \\\hline
&  & \\[-6pt]%
Autocorrelator & $\exp\left[  -\lambda_{c}t-\beta^{2}t/\ln^{2}\!t\right]  $,
Eq. (\ref{400}) & $p_{0}^{\left(  \mathrm{eq}\right)  }+\mathrm{const~}%
t^{\eta}\exp\left[  -3\left(  \frac{a}{2}\right)  ^{2/3}t^{1/3}\right]  $,
Eqs.$~$(\ref{480}), (\ref{550})\\[-6pt]
&  & \\\hline
&  & \\[-6pt]%
Propagator at $l\sim t$ & $\frac{1}{\sqrt{2\pi Dt}}\exp\left[  -\frac{\left(
l-vt\right)  ^{2}}{2Dt}\right]  $, Eq. (\ref{450}) & $\frac{1}{\sqrt{2\pi Dt}%
}\exp\left[  -\frac{\left(  l-vt\right)  ^{2}}{2Dt}\right]  $, Eqs.
(\ref{450}), (\ref{570})\\[-6pt]
&  & \\\hline
&  & \\[-6pt]%
Propagator at $t{\rightarrow}\infty\phantom{o}$ & $\mu_{0}^{l}\left(
-\lambda_{c}\right)  \exp\left[  -\lambda_{c}t-\beta^{2}t/\ln^{2}\!t\right]  $
& $\phantom{o}p_{l}^{\left(  \mathrm{eq}\right)  }+ct^{-\zeta}\exp\left[
-3\left(  \frac{a}{2}\right)  ^{2/3}\left(  t-l/v\right)  ^{1/3}\right]  ,$
Eqs. (\ref{486}), (\ref{580})$\phantom{o}$\\[-6pt]
&  & \\\hline
\end{tabular}
\label{results}
\end{table*}%

In this paper we study properties of the Laplacian operator
\begin{equation}
L_{ij}=q_{i}\delta_{ij}-A_{ij} \label{11}%
\end{equation}
on an uncorrelated random network near the lower edge of its spectrum, and,
respectively, the hopping motion of some carrier (\textquotedblleft
signal\textquotedblright) from one vertex to another at large times. This
operator corresponds to the process described the following dynamic equations
for the probability $p_{ij}(t)$ that at time $t$ a particle is at vertex $i$
if at time $0$ it was at vertex $j$:
\begin{equation}
\dot{p}_{ij}\left(  t\right)  =\sum_{k=1}^{N}A_{ik}p_{kj}(t)-q_{i}%
p_{ij}(t),\ p_{ij}(0)=\delta_{ij}. \label{13}%
\end{equation}
This is a random walk where the rate of hopping along any edge is set to one.
Other versions of the Laplace operator and corresponding processes, which are
also widely discussed in literature, are listed in Appendix \ref{a10000}.

We use the configuration model of an uncorrelated network
\cite{b80,bc78}, which is a maximally random network with a given
degree distribution. It is convenient that (i) this model is statistically
homogeneous, (ii) all its vertices are statistically independent, and (iii) it
has a locally tree-like structure. We consider only infinite networks, that
is, first we tend the total number of vertices $N$ to infinity (\emph{the
thermodynamic limit}) and only afterwards study network characteristics. If,
say, we study a random walk, then a particle should be still much closer to an
initial vertex than the diameter of the network $\sim\ln N$. In other words,
we consider the process at so short times that the number of vertices, where
the walking particle may be found, is negligible compared with the network's
size $N$. We will see that this imposes strong limitations to the
applicability of our results due to the ``small world'' feature of the networks under consideration.

\begin{figure}[t]
\begin{center}
\scalebox{0.73}{\includegraphics{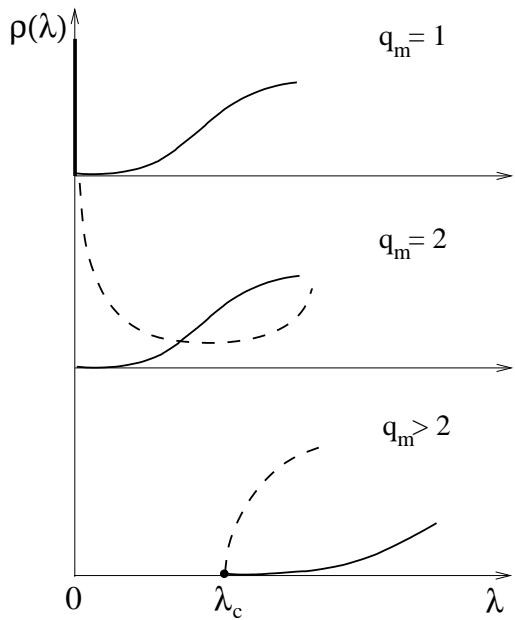}}
\end{center}
\caption{Laplacian spectral density $\rho\left(  \lambda\right)  $ for
networks with different minimum vertex degree $q_{m}$: (a) $q_{m}>2$, (b)
$q_{m}=2$, and $q_{m}=1$ (the vertical line at $\lambda=0$ represents a 
$\delta$-function peak).}%
\label{spectra}%
\end{figure}

We will show that for the Laplacian spectra and for random walks, the crucial
property of the random uncorrelated network is the minimum degree $q_{m}$ of
its vertices. We suppose that the value of the degree distribution at $q_{m}$
essentially differs from $0$ and $1$. We also assume that $q_{m}>0$, because
the contribution of isolated vertices is trivial. Our results are summarized
in Table~\ref{results} and in Fig.~\ref{spectra}. Note an unusual singularity of the spectral density in the case $q_m>2$. 

As is natural, the calculation of the spectrum is reduced to the study 
of the trace of the Laplace operator's resolvent. To describe the propagation of
the signal in the network, one must know the non-diagonal elements
of the resolvent. Here we calculate the asymptotics of their average
values. It allows us to obtain the time and distance dependences of the signal's
\emph{propagator} $\bar{p}_{ij}\left(  t\right)  =\bar{p}_{l}\left(  t\right)
$ 
when the distance between initial $i$ and final $j$ vertices, 
$l$ is much smaller than the diameter of the network, $\bar{l}\sim\ln N$. 

Why is the minimum vertex degree so important in these problems? 
Note that in respect of random walks and Laplacian operator related problems, infinite uncorrelated networks are equivalent to infinite Bethe lattices with coinciding degree distributions.  
(Recall that a Bethe lattice is an infinite tree without borders.) 
Let us compare two Bethe lattices---random, with the minimum coordination number $q_m$, and regular, with the coordination number equal to $q_m$.  
It is clear that the 
autocorrelator in the random Bethe lattice cannot decay slower than in the regular Bethe lattice with this $q_{m}$. 
If $q_m>2$, then in this regular Bethe lattice, $\bar{p}_{ii}\left(  t\right)  =\bar{p}%
_{0}\left(  t\right)  \sim$ $t^{-3/2}\exp\left(  -\lambda_{c}t\right)  $, where
\begin{equation}
\lambda_{c}=q_{m}-2\sqrt{q_{m}-1}. \label{15}%
\end{equation} 
$\lambda_{c}$ is also the spectral boundary in the Laplacian eigenvalue density $\rho(\lambda)$ of this regular Bethe lattice, where $\rho(\lambda)\sim\sqrt{\lambda-\lambda_{c}}$, near $\lambda_{c}$.  
Thus, the spectral boundary for an infinite uncorrelated network in principle cannot be lower than that for the regular Bethe lattice with the same $q_m$. 
Moreover, these borders coincide. The reason for this is the following feature of the configuration model of an uncorrelated network. 
Let the number of vertices $N$ in this model approach infinity. Then the mean number of given finite regular subgraphs with coordination number $q_m$ grows proportionally to $N$. We stress that although this number rapidly decreases with a size of these subgraphs, it is proportional to $N$ for any given subgraph size.  In the arbitrarily large subgraphs, the lowest eigenvalues are arbitrarily close to the spectral boundary of the corresponding regular Bethe lattice. 
The number of these eigenvalues is proportional to the number of these subgraphs and so proportional to $N$.  
Now recall that the total number of eigenvalues in the spectrum is $N$. Therefore, indeed, the spectral borders for the configuration model and for the regular Bethe lattice with $q_m$ coincide. 

The statistics of these regular tree subgraphs determine the singularity of the resulting $\rho(\lambda)$ at the edge $\lambda_c$. The rapid decrease of of the number of these subgraphs with their size results in specific singularities, with all derivatives zero, represented in Table~\ref{results}. 

The random networks with $q_{m}=1,2$ markedly differ from those with $q_m>2$. 
In the configuration model with $q_{m}=1,2$, chains and chain-like subgraphs are statistically essential. Let us first discuss the case $q_m=2$. 
The Bethe lattice with coordination number $2$ is a usual infinite chain. It has the spectral boundary $\lambda_{c}=0$. Near this edge, $\rho\left(  \lambda\right)  \sim\lambda^{-1/2}$. Thus, the edge of the spectrum of the uncorrelated network with $q_m=2$ is zero.  
We will show that the statistics of chain subgraphs in this configuration model differ from those for the case $q_m>2$. This results in different asymptotics presented in Table~\ref{results} and, schematically, in Fig.~\ref{spectra}.

\begin{figure}
\begin{center}
\scalebox{0.4}{\includegraphics{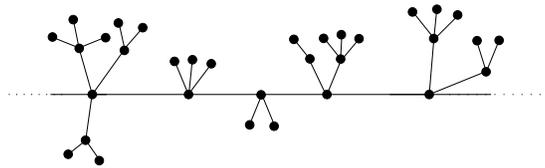}}
\end{center}
\caption{
Chain with finite tree-like branches in a random network with the minimum vertex degree $q_m=1$.}%
\label{branches}%
\end{figure}

If $q_{m}=1$, chain-like subgraphs are also present in the configuration model. These are, however, more chains (see Fig.\ref{branches}) with branches attached. Nonetheless, these subgraphs result in the spectrum edge $\lambda_c=0$ and in the same asymptotics as for $q_m=2$. When $q_{m}=1$, numerous finite components components are present in the network. Their mean number is proportional to $N$. Each of connected components gives one zero eigenvalue in the spectrum. This leads to a $\delta$-function peak at $\lambda=0$ in the
spectral density.     

The found singularities of $\rho(\lambda)$, with all derivatives zero, have a direct consequence for observations in finite networks. Even in a huge uncorrelated network, the observed minimum eigenvalue $\lambda_2$ will be far from the spectral edge $\lambda_c$ predicted for an infinite network. 
Let us roughly estimate $\lambda_2(N)$ based on the spectral densities from Table~\ref{results}. 
The condition $N\int_{\lambda_c}^{\lambda_2(N)}d\lambda \rho(\lambda) \sim 1$ leads to the following dependences in the range of large $N$. If $q_m=1,2$, then 
\begin{equation}
\lambda_2(N) \sim (\ln N)^{-2}
, 
\label{17}%
\end{equation} 
and if $q_m>2$, then 
\begin{equation}
\lambda_2(N)-\lambda_c \sim (\ln\ln N)^{-2}
.
\label{19}%
\end{equation} 
Thus the approach of $\lambda_2(N)$ to $\lambda_c$ is extremely slow. 
Note that a very slow convergence of $\lambda_2$ was recently observed in  
the numerical work of Kim and Motter, Ref.~\cite{km07}, in which $\lambda_2$ and $\lambda_c$ were compared for networks up to 4\,000 vertices.   

In Sec.~\ref{formulation} we strictly formulate the problem. In
Sec.~\ref{main} we derive a basic set of integral equations. Solving these
equations enables us to obtain the Laplacian spectrum $\rho\left(
\lambda\right)  $ for uncorrelated random networks and to describe the random
walk on the networks in the thermodynamic limit. In Sec.~\ref{finite} we study
the final value of the propagator $\bar{p}_{i}^{\left(  \mathrm{eq}\right)
}=\bar{p}_{l}\left(  t\rightarrow\infty\right)  $, which is the equilibrium
probability to find a signal at distance $l$ from a starting vertex. We
describe $\bar{p}_{i}^{\left(  \mathrm{eq}\right)  }$ in terms of $l$ and of
the degree distribution $\Pi\left(  q\right)  $. Furthermore, we find the
coefficient of  the $\delta\left(  \lambda\right)  $ term. In
Sec.~\ref{allqm} we present general solutions of the integral equations of
Sec.~\ref{main} and analyse them in three distinct cases: $q_{m}>2$, $q_{m}%
=2$, and $q_{m}=1$. In Sec.~\ref{conclusions} we summarize our results and
methods and discuss conditions for their applicability. Technical details are
given in Appendices.

\section{Formulation of the problem}

\label{formulation}

The problem of the Laplacian spectrum of a random network is completely
equivalent to that of the time dependence of the averaged
\emph{autocorrelator} $\bar{P}_{0}\left(  t\right)  =\left\langle
P_{ii}\left(  t\right)  \right\rangle \,$ for a random walk. This autocorrelator is the
probability that a particle returns to the starting vertex after a time $t$.
This quantity is related to the eigenvalue density%
\begin{equation}
\rho\left(  \lambda\right)  =\frac{1}{N}\left\langle \sum_{n=1}^{N}%
\delta\left(  \lambda-\lambda_{n}\right)  \right\rangle \label{20}%
\end{equation}
in the following way:
\begin{equation}
\bar{P}_{0}\left(  t\right)  =\int_{0}^{\infty}d\lambda~e^{-\lambda t}%
\rho\left(  \lambda\right)  , \label{30}%
\end{equation}
where $\lambda_{k}$ are (nonnegative) eigenvalues of the Laplace operator on
the network:%
\begin{equation}
\hat{L}\mathbf{a}^{\left(  k\right)  }=\lambda_{k}\mathbf{a}^{\left(
k\right)  }\,,\ \ \mathbf{a}^{\left(  k\right)  }=\left(  a_{1}^{\left(
k\right)  },a_{2}^{\left(  k\right)  },\dots,a_{N}^{\left(  k\right)
}\right)  , \label{40}%
\end{equation}%
\begin{equation}
\left(  \hat{L}\mathbf{x}\right)  _{i}\mathbf{=}\sum_{j}A_{ij}\left(
x_{i}-x_{j}\right)  =q_{i}x_{i}-\sum_{j}A_{ij}x_{j}. \label{50}%
\end{equation}

We assume, that a particle moves from vertex to vertex by hopping along edges.
To every edge we ascribe a \emph{hopping rate} $w_{ij}$, which is the probability to
move from vertex $j$ to vertex $i$ per unit time. Hopping rates are assumed to
be symmetric and equal $1$ for every edge, $w_{ij}=w_{ji}=A_{ij}=0$ or $1$. In this paper we fix $w_{ij}$ but not the escape rate of a particle
from a vertex, see Appendix \ref{a10000} where other forms of a Laplace operator are listed. 
It turns out that our main conclusions are also valid if the escape rate from a vertex is fixed. This case will be discussed in detail in our next works. 
Assume that at $t=0$ the particle is
at vertex $j$. Its motion is governed by the master equation for the
propagator, which is the probability $p_{ij}\left(  t\right)  $ that at time
$t$ the particle is at vertex $i$,%
\begin{equation}
\dot{p}_{ij}\left(  t\right)  =\sum_{k=1}^{N}\left[  w_{ik}p_{kj}\left(
t\right)  -w_{ki}p_{ij}\left(  t\right)  \right]  =\sum_{j=1}^{N}A_{ij}\left[
p_{ij}\left(  t\right)  -p_{ji}\left(  t\right)  \right]  =\sum_{k=1}%
^{N}A_{ik}p_{kj}\left(  t\right)  -q_{i}p_{ij}\left(  t\right).  
\label{70}%
\end{equation}
This equation is supplied 
with the initial condition $p_{ij}\left(  0\right)  =\delta_{ij}$. What is the
value of the probability
\begin{equation}
\bar{p}_{n}\left(  t\right)  =\frac{1}{N}\left\langle \sum_{d\left(
i,j\right)  =l}p_{ij}\left(  t\right)  \right\rangle \label{60}%
\end{equation}
that at time $t$ the particle is at distance $d\left(  i,j\right)  =n$ from
a starting vertex? (The distance is the minimum shortest path between two
vertices.) Here $\left\langle \cdots\right\rangle $ means the average over
some statistical ensemble of graphs (over that of the configuration model in
our case).

In the Laplace representation,
\begin{equation}
P_{ik}\left(  s\right)  =\int_{0}^{\infty}dt~p_{ik}\left(  t\right)  e^{-st},
\label{65}%
\end{equation}
the propagator is the resolvent of the Laplace operator:%
\begin{equation}
\hat{P}\left(  s\right)  =\left(  s+\hat{L}\right)  ^{-1}. \label{80}%
\end{equation}
Consequently, the density of eigenvalues is expressed in terms of the analytic
continuations of the averaged values of the autocorrelator:%
\begin{equation}
\rho\left(  \lambda\right)  =\frac{1}{2\pi i}\left[  \bar{P}_{0}\left(
-\lambda-i0\right)  -\bar{P}_{0}\left(  -\lambda+i0\right)  \right]  .
\label{90}%
\end{equation}
The inverse relation is 
\begin{equation}
\bar{p}_{0}\left(  t\right)  =\int_{-i\infty+\delta}^{+i\infty+\delta}%
\frac{ds}{2\pi i}e^{st}\bar{P}_{0}\left(  s\right)  =\int_{0}^{\infty}%
d\lambda~e^{-\lambda t}\rho\left(  \lambda\right)  . \label{100}%
\end{equation}

\section{Main equations}

\label{main}

We assume the thermodynamic limit: $N\rightarrow\infty$, and the fraction of
vertices with a degree $q$, $N\left(  q\right)  /N\rightarrow\Pi\left(
q\right)  $. Here $\Pi\left(  q\right)  $ is a given degree distribution with
a finite second moment, $\sum_{q}q^{2}\Pi\left(  q\right)  <\infty$. In this
limit, almost all finite subgraphs are trees, i.e., they have no closed loops
within. The network is uncorrelated, i.e., degrees of any pair of vertices,
connected or not, are independently distributed random variables. These
features allowed us to describe the statistics of intertvertex distances
\cite{dms03}. The problem under consideration is actually related to that work.

The equation for the resolvent of the Laplace operator (\ref{80}) is
\begin{equation}
sP_{ik}=\delta_{ik}+\sum_{j}A_{ij}\left(  P_{jk}-P_{ik}\right)  =\delta
_{ik}+\sum_{j}A_{ij}P_{jk}-q_{i}P_{ik}. \label{110}%
\end{equation}
Without lack of generality we choose the initial vertex $j=0$.

By definition, the $n$-th connected component of a vertex $i$ is a subgraph,
containing all vertices $j$ within the distance $d\left(  i,j\right)  \leq n$
from the vertex $i$. For any finite $n$, in an infinite graph almost any
$n$-th connected component of vertex $0$ is a tree. Actually, we analyse a
random Bethe lattice. Degrees of its vertices are independent random
variables. Its arbitrary chosen central vertex has the vertex distribution
function $\Pi\left(  q\right)  =\left\langle \delta\left(  q_{0}-q\right)
\right\rangle $. The other vertices have degree distributions equal among
themselves but different from $\Pi\left(  q\right)  .$ Non-central vertex $i$
of a degree $q_{i}$ has one edge directed to the central vertex and
$b_{i}=q_{i}-1\geq0$ edges directed from it. Here $b$ is the branching number
of the vertex. Its distribution is given by
\begin{equation}
\Pi_{1}\left(  b\right)  =\frac{1}{2L}\sum_{i,j=1}^{N}\left\langle
A_{ij}\delta\left(  b_{j}-b\right)  \right\rangle =\frac{\left(  b+1\right)
}{\bar{q}}\Pi\left(  b+1\right)  , \label{115}%
\end{equation}
where $\bar{q}=2L/N$.
\begin{equation}
L=\frac{1}{2}\sum_{i,j=1}^{N}A_{ij}=\frac{1}{2}\sum_{i=1}^{N}q_{i}
\label{117}%
\end{equation}
is the total number of edges in the graph. In Eq.~(\ref{115}) we assume that
$N\rightarrow\infty$, $L\rightarrow\infty$, but $2L/N\rightarrow\bar{q}%
=\sum_{q}q\Pi\left(  q\right)  $, where $\bar{q}$ is some finite number.
$\Pi_{1}\left(  b\right)  $ is the probability that a randomly chosen end
vertex of a randomly chosen edge in the graph has $b=q-1$ edges apart from the
chosen edge itself. It is convenient to use distributions $\Pi$ and $\Pi_{1}$
in Z-representation (see Appendix \ref{Z}).

\begin{figure}[t]
\begin{center}
\scalebox{0.41}{\includegraphics{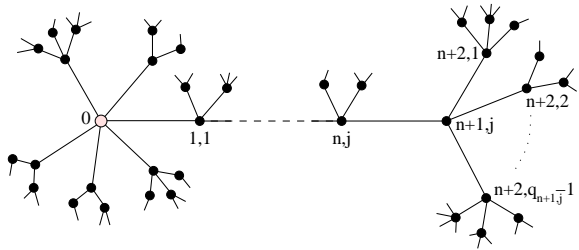}}
\end{center}
\par
\caption{ Vertex $\left(  n+1,i\right)  $ at distance $n$ from the starting
vertex $0$, its ``ancestor'' $\left(  n,j\right)  $, and its ``descendants''
$\left(  n+2,k\right)  $. $k=1,2,\dots q_{n+1,k}-1$, $q_{n+1,k}$ is the degree
of the vertex $\left(  n+1,i\right)  $. }%
\label{tree}%
\end{figure}

Let $\left(  n,i\right)  $ and $\left(  n+1,j\right)  $ be two vertices
connected by an edge and at the distances $n-1$ and $n$, respectively, from
the starting vertex $0$. We introduce the following random variable (see
Fig.~\ref{tree}):%
\begin{equation}
\tau_{n,ij}\left(  s\right)  =\frac{P_{n,i}\left(  s\right)  -P_{n+1,j}\left(
s\right)  }{P_{n,i}\left(  s\right)  }. \label{120}%
\end{equation}

It is obvious that the statistical properties of this variable are independent
of the particular choice of vertex $i$ in the $n$-th shell of the initial
vertex $0$. The graph ensemble under consideration is completely defined by
the degree distribution. All graphs with a given degree distribution have the
same statistical weights. This, in particular, implies \emph{the statistical
homogeneity} of the ensemble. First, we randomly choose vertex $0$. Second, we
label all other vertices by two indices: the first one is the distance from
vertex $0$ (the shell's number), and the second index labels vertices within
the shell. Third, we consider $P_{ni}$, which is the matrix element of the
resolvent for the pair---vertex $0$ and vertex $i$ at distance $n$ from vertex
$0$. It is a fluctuating random variable but its statistical properties are
independent of the choice of $i$, because every averaging includes averaging
over all vertices in shell $n$. The other fluctuating quantity in
Eq.~(\ref{120}), $P_{n+1,j}$, is, of course, correlated with $P_{n,i}$.
Nonetheless, due to \emph{the statistical independence} of the vertices, this
correlation is independent of the particular choice of the connected pair of
vertices. Therefore, we can define the distribution function of $\tau_{n,ij}$,
which is independent of $i,j$. In the Laplace representation this distribution is defined as 
\begin{equation}
T_{n}\left(  s,x\right)  =\left\langle \exp\left[  -x\tau_{n,ij}\left(
s\right)  \right]  \right\rangle . \label{130}%
\end{equation}

Now let us recall that in the infinite network all finite connected components
are trees. Moreover, in the thermodynamic limit the statistical properties of
all $\tau$ variables are the same, i.e., they are independent of $n$ too (see
more detailed discussion in Appendix \ref{T}). It implies the following
important consequences. (i) In the thermodynamic limit, i.e., for an infinite
network, $T_{n}\equiv T$ is independent of $n$. (ii) It is possible to
obtain the closed equation for $T\left(  s,x\right)  $. (iii) The density of
eigenvalues $\rho\left(  \lambda\right)  $, and, consequently, the
autocorrelator $\bar{p}_{0}\left(  t\right)  $ can be expressed in terms of
$T\left(  s,x\right)  $ (see Appendix \ref{T}).

Equation for $T\left(  s,x\right)  $ may be written as
\begin{equation}
e^{x}T\left(  s,x\right)  =1+\sqrt{x}\int_{0}^{\infty}\frac{dy}{\sqrt{y}}%
I_{1}\left(  2\sqrt{xy}\right)  e^{-\left(  1+s\right)  x}\varphi_{1}\left[
T\left(  s,y\right)  \right]  , \label{140}%
\end{equation}
where $I_{1}$ is a modified Bessel function, and $\varphi_{1}\left(  z\right)
$ is the degree distribution of non-central vertices branching numbers in
Z-representation (see Appendix \ref{Z}). As $\operatorname{Re}s>0$, this
function has a solution with all properties of the Laplace transform of the
distribution density of a nonnegative random variable. This statement may be
proved by using an approach of Ref.~\cite{gk83}. 

The function $\varphi_{1}(z)$ has the following properties: (i) $\varphi
_{1}\left(  0\right)  =\Pi_{1}\left(  1\right)  $, which is the concentration
of vertices with degree $1$ (\textquotedblleft dead ends\textquotedblright),
and (ii) if the degree distribution $\Pi_{1}\left(  q\right)  $ decays faster
than any exponent for $q\rightarrow\infty$, then $z=1$ is a point of
singularity of $\varphi_{1}\left(  z\right)  $. The function $\varphi
_{1}\left(  z\right)  $ in the complex plane is analytic within the circle
$\left\vert z\right\vert <1$, $\varphi_{1}\left(  1\right)  =1$. The parameter
$\varphi_{1}\left(  0\right)  $ is the crucial one in the division of the
graph into connected components, see Appendix~\ref{Z}. We show in this
Appendix that the autocorrelator $\bar{P}_{0}\left(  s\right)  $ in the
Laplace representation is given by
\begin{equation}
\bar{P}_{0}\left(  s\right)  =\int_{0}^{\infty}dx~e^{-sx}\varphi\left[
T\left(  s,x\right)  \right]  , \label{150}%
\end{equation}
where $\varphi(z)$ is the Z-transformation of the degree distribution $\Pi
(q)$. The functions $\varphi$ and $\varphi_{1}$ are connected as $\varphi
_{1}\left(  z\right)  =\varphi'\left(  z\right)  /\varphi'\left(  1\right)
$, so that $\varphi\left(  1\right)  =\varphi_{1}\left(  1\right)  =1$. The
density of eigenvalues $\rho\left(  \lambda\right)  $ and time-dependent
autocorrelator $\bar{p}_{0}\left(  t\right)  $ can be obtained from Eqs.
(\ref{90}) and (\ref{100}), respectively.

The propagator $\bar{p}_{n}\left(  t\right)  $ at $n>0$ may be expressed in
terms of some functions $U_{n}\left(  s,x\right)  $, for which we have a
linear recursion, relating $U_{n}$ to $U_{n-1}$ (see Appendix \ref{U}). These
functions are introduced in the following way. Let us choose two vertices
$\left(  n,i\right)  $ and $\left(  n+1,j\right)  $, connected by an edge
(Fig.~\ref{tree}). We define $S_{n;i,j}^{\left(  l\right)  }\left(  s\right)
$ as
\begin{equation}
S_{n;i,j}^{\left(  l\right)  }\left(  s\right)  =\frac{1}{P_{n,i}\left(
s\right)  }\sum_{\left\{  k\right\}  }P_{n+l,k}\left(  s\right)  . \label{160}%
\end{equation}
Here the summation is over all those vertices at a distance $n+l$ from vertex
$0$, whose shortest path to vertex $0$ runs along the edge $\left(
n,i\right)  \rightarrow\left(  n+1,j\right)  $. In other words, the sum in
Eq.~(\ref{160}) is over all vertices of the $l$-th generation of the branch
beginning from a chosen edge. For example, $S_{n,;i,j}^{\left(  1\right)
}\left(  s\right)  =\sum_{k=1}^{b_{n+1,j}}P_{n+1,k}\left(  s\right)  $, as one
can see from Fig.~\ref{tree}. Due to the statistical homogeneity of the
network ensemble the statistical properties of random variables $S_{n,;i,j}%
^{\left(  l\right)  }$ are independent of the choice of vertices $i$ and $j$,
if they are connected by an edge. For an infinite network, this statistics is
also independent of $n$. The recursion relation can be derived for the
following averaged quantity which depends only on $l$:%
\begin{equation}
U_{l}\left(  s,x\right)  =\left\langle S_{n;i,j}^{\left(  l\right)  }\left(
s\right)  \exp\left[  -x\tau_{n,ij}\left(  s\right)  \right]  \right\rangle .
\label{170}%
\end{equation}
The recursion relation is derived in Appendix~\ref{U}. It is of the following
form:%
\begin{equation}
U_{l}\left(  s,x\right)  =e^{-x}\int_{0}^{\infty}dy~I_{0}\left(  2\sqrt
{xy}\right)  e^{-\left(  1+s\right)  y}\varphi_{1}^{\prime}\left[  T\left(
s,y\right)  \right]  U_{l-1}\left(  s,y\right)  , \label{180}%
\end{equation}
where $I_{0}$ is a modified Bessel function of zero order. This recursive
relation is supplied with the initial condition:%
\begin{equation}
U_{1}\left(  s,x\right)  =\left(  1+\frac{\partial}{\partial x}\right)
T\left(  s,x\right)  . \label{190}%
\end{equation}
Finally, the Laplace-transformed propagator $\bar{P}_{l}\left(  s\right)  $ is
expressed as
\begin{equation}
\bar{P}_{l}\left(  s\right)  =\int_{0}^{\infty}dx~e^{-sx}\varphi^{\prime
}\left[  T\left(  s,x\right)  \right]  U_{l}\left(  s,x\right)  . \label{200}%
\end{equation}

Equation~(\ref{180}) may be presented in the form:%
\begin{equation}
U_{n}=\widehat{M}U_{n-1}, \label{2010}%
\end{equation}
where $\widehat{M}$ is a linear integral operator. Let $\mu_{m}^{-1}\left(
s\right)  $ and $\psi_{m}\left(  s,x\right)  $ be its eigenvalues and
eigenfunctions, respectively:%
\begin{equation}
\mu_{m}\left(  s\right)  e^{-x}\int_{0}^{\infty}dy~I_{0}\left(  2\sqrt
{xy}\right)  e^{-\left(  1+s\right)  y}\varphi_{1}^{\prime}\left[  T\left(
s,y\right)  \right]  \psi_{m}\left(  s,y\right)  =\psi_{m}\left(  s,x\right)
. \label{2020}%
\end{equation}
Note that the general theory of integral operators is usually formulated in
terms of $\mu_{m}^{-1}$, which are called their eigenvalues. Operator
$\widehat{M}$ becomes Hermitian after the substitution $\psi_{m}\left(
x\right)  =e^{sx/2}\left\{  \varphi_{1}^{\prime}\left[  T\left(  s,x\right)
\right]  \right\}  ^{-1/2}\chi\left(  s,x\right)  $. The kernel of the
integral operator in Eq. (\ref{2020}) is bounded \cite{zkkmrs68} if
\begin{equation}
\int_{0}^{\infty}dx\int_{0}^{\infty}dyI_{0}^{2}\left(  2\sqrt{xy}\right)
e^{-\left(  2+s\right)  \left(  x+y\right)  }\varphi_{1}^{\prime}\left[
T\left(  s,x\right)  \right]  \varphi_{1}^{\prime}\left[  T\left(  s,y\right)
\right]  <\infty. \label{2025}%
\end{equation}
This condition is always satisfied if $s>0$. Therefore, 
according to theorems about integral equations with a
Hermitian bounded kernel \cite{zkkmrs68}, all eigenvalues of this operator are real and finitely degenerate.  
They
form a discrete sequence bounded from below, without any condensation point
except $\mu=\infty$. Eigenfunctions are orthogonal and normalizable with the
weight function $e^{-sx}\varphi_{1}^{\prime}\left[  T\left(  s,x\right)
\right]  $:%
\begin{equation}
\int_{0}^{\infty}dx~\varphi_{1}^{\prime}\left[  T\left(  s,x\right)  \right]
e^{-sx}\psi_{k}\left(  s,x\right)  \psi_{m}\left(  s,x\right)  =\delta_{km}.
\label{2030}%
\end{equation}
Hence the solution of the recursive relation (\ref{180}) may be presented as a
series in the complete orthonormal set $\left\{  \psi_{m}\right\}  $,%
\begin{equation}
U_{l}\left(  s,x\right)  =\sum_{m}A_{m}\left(  s\right)  \mu_{m}^{1-l}\left(
s\right)  \psi_{m}\left(  s,x\right)  . \label{2040}%
\end{equation}
Taking into account the initial condition (\ref{190}) and the orthonormality
condition (\ref{2030}), the coefficients in this series may be written as
\begin{equation}
A_{m}\left(  s\right)  =\int_{0}^{\infty}dx~\psi_{m}\left(  x\right)
\varphi_{1}^{\prime}\left[  T\left(  s,x\right)  \right]  e^{-sx}\left(
1+\frac{\partial}{\partial x}\right)  T\left(  s,x\right)  . \label{2050}%
\end{equation}
Substituting Eq.~(\ref{2040}) into Eq.~(\ref{200}), we get for $l>0$ the
following relation:%
\begin{equation}
\bar{P}_{l}\left(  s\right)  =\sum_{m}A_{m}\left(  s\right)  B_{m}\left(
s\right)  \mu_{m}^{1-l}\left(  s\right)  , \label{2060}%
\end{equation}
where
\begin{equation}
B_{m}\left(  s\right)  =\int_{0}^{\infty}dx~e^{-sx}\varphi^{\prime}\left[
T\left(  s,x\right)  \right]  \psi_{m}\left(  s,x\right)  . \label{2070}%
\end{equation}

The resulting propagator $\bar{P}_{l}\left(  s\right)  $ satisfies the
condition of the conservation of the number of particles/signals, which in the
Laplace representation is $\sum_{l=0}^{\infty}\bar{P}_{l}\left(  s\right)
=1/s$. Taking into account Eq. (\ref{2060}) gives the following form of this condition:  
\begin{equation}
\bar{P}_{0}\left(  s\right)  +\sum_{m}\frac{A_{m}\left(  s\right)
B_{m}\left(  s\right)  }{\mu_{m}\left(  s\right)  -1}=\frac{1}{s}.
\label{2080}%
\end{equation}

\section{Contribution of finite connected components}

\label{finite}

When the minimum vertex degree in the uncorrelated network $q_{m}\geq2$, then
(in the thermodynamic limit) 
the network consists of one connected
component. If, however, $q_{m}=1$, i.e., $\Pi\left(  1\right)  =\Pi_{1}\left(
1\right)  =\varphi^{\prime}\left(  0\right)  =\bar{q}\varphi_{1}\left(
0\right)  \neq0$, then connected components exist even in the thermodynamic
limit. Their contribution to the propagator at $t\rightarrow\infty$ is
obvious, and can be calculated in a straightforward way. We, however, find
this contribution by using the technique described in 
Sec.~\ref{main} for the sake of illustration.

We set in $T\left(  s,x\right)  $ the limit $s\rightarrow0$ and $x\rightarrow
\infty$, with $sx$ fixed, assuming that there exists a limiting function
\begin{equation}
\Theta\left(  z\right)  =\lim_{s\rightarrow\infty}T\left(  s,z/s\right)  .
\label{210}%
\end{equation}
In Appendix \ref{A1} we derive the following equation for $\Theta$:
\begin{equation}
\Theta\left(  z\right)  =e^{-z}\varphi_{1}\left[  \Theta\left(  z\right)
\right]  . \label{220}%
\end{equation}
Comparing this equation with Eq.~(\ref{Z50}) from Appendix \ref{Z}, one can
conclude that $\Theta\left(  z\right)  =H\left(  e^{-z}\right)  $, that is
$\varphi\left[  \Theta\left(  z\right)  \right]  =\varphi\left[  H\left(
e^{-z}\right)  \right]  =\left\langle \exp\left(  -zM_{i}\right)
\right\rangle $. Here $M_{i}$ is the size of the connected component with a
randomly chosen vertex $i$. It is obvious that the giant connected component,
whose size is $\sim N$, does not contribute to $\Theta\left(  z\right)  $ at
any $z>0$ as $N\rightarrow\infty$. From Eqs. (\ref{150}) and (\ref{210}), we
obtain a clear result for the limiting value of the autocorrelator,
\begin{equation}
p_{0}^{\left(  \mathrm{eq}\right)  }\equiv\bar{p}_{0}\left(  t=\infty\right)
=\lim_{s\rightarrow0}s\bar{P}_{0}\left(  s\right)  =\int_{0}^{\infty}%
dz\varphi\left[  \Theta\left(  z\right)  \right]  =\left\langle \frac{1}%
{M_{i}}\right\rangle . \label{230}%
\end{equation}
It means that the equilibrium distribution of the signal is homogeneous within
its connected component. Passing from the variable $z$ to $x=\Theta\left(
z\right)  $ and using Eq.~(\ref{220}), we calculate this integral:%
\begin{equation}
p_{0}^{\left(  \mathrm{eq}\right)  }=\varphi\left(  t_{c}\right)  -\frac{1}%
{2}t_{c}\varphi^{\prime}\left(  t_{c}\right)  . \label{235}%
\end{equation}
This result also has a different meaning,%
\begin{equation}
p_{0}^{\left(  \mathrm{eq}\right)  }=\frac{1}{N}\left\langle \sum_{i=1}%
^{N}\frac{1}{M_{i}}\right\rangle =\frac{1}{N}\left\langle \sum
_{\mathrm{Clusters}}1\right\rangle =\frac{N_{c}}{N}, \label{240}%
\end{equation}
where $N_{c}$ is the total number of finite connected components.

A nonzero equilibrium value of the autocorrelator indicates that the
degeneracy of the Laplacian eigenvalue $\lambda=0$ is $\sim N$. The
eigenvectors of this eigenvalue may be chosen in the following way. Each such
eigenvector has unit vector components in one connected component and zeros in
all others. The degeneracy is equal to the total number of connected component
in the network.

The $t\rightarrow\infty$ contribution of finite connected components to
$\bar{p}_{l}\left(  t\right)  $ at $l>0$ may be extracted from the functions:%
\begin{equation}
u_{l}\left(  z\right)  =\lim_{s\rightarrow\infty}U_{l}\left(  s,z/s\right)  .
\label{250}%
\end{equation}
In this limit the recurrent relation (\ref{180}) turns into
\begin{equation}
u_{l}\left(  z\right)  =e^{-z}\varphi_{1}^{\prime}\left[  \Theta\left(
z\right)  \right]  u_{l-1}\left(  z\right)  \label{260}%
\end{equation}
(see derivation in Appendix \ref{A1}). From Eq. (\ref{190}) it also follows
that $u_{1}\left(  z\right)  =\Theta\left(  z\right)  $, so
\begin{equation}
u_{l}\left(  z\right)  =\left\{  e^{-z}\varphi_{1}^{\prime}\left[
\Theta\left(  z\right)  \right]  \right\}  ^{l-1}\Theta\left(  z\right)  .
\label{270}%
\end{equation}

Let us calculate $p_{l}^{\left(  \mathrm{eq}\right)  }=\bar{p}_{l}\left(
t=\infty\right)  $. The stationary value of $P_{ij}\left(  t\right)  $ at
$t\rightarrow\infty$ is equal to $1/M_{j}$, where $M_{j}$ is connected
component with an initial vertex $j$. Consequently,
\begin{equation}
p_{l}^{\left(  \mathrm{eq}\right)  }=\left\langle \frac{Q_{j}^{\left(
l\right)  }}{M_{j}}\right\rangle , \label{280}%
\end{equation}
where $Q_{j}^{\left(  l\right)  }$ is the number of vertices at distance $l$
from vertex $j$. Using Eqs. (\ref{200}), (\ref{250}) and (\ref{270}), we get
\begin{equation}
p_{l}^{\left(  \mathrm{eq}\right)  }=\int_{0}^{\infty}dx~e^{-x}\Theta\left(
x\right)  \varphi^{\prime}\left[  \Theta\left(  x\right)  \right]  \left\{
e^{-x}\varphi_{1}^{\prime}\left[  \Theta\left(  x\right)  \right]  \right\}
^{l-1}. \label{290}%
\end{equation}
At large $l$, the region $z\ll1$, where $\Theta\left(  z\right)  $ is close to $t_{c}$, gives the main contribution to the integral in Eq. (\ref{290}). As a result, at
large $l$, we have
\begin{equation}
p_{l}^{\left(  \mathrm{eq}\right)  }\approx\frac{b}{n}\left[  \varphi
_{1}^{\prime}\left(  t_{c}\right)  \right]  ^{n},\ \ b=\frac{\bar{q}t_{c}%
^{2}\left[  1-\varphi_{1}^{\prime}\left(  t_{c}\right)  \right]  }{\varphi
_{1}^{\prime}\left(  t_{c}\right)  \left[  1-\varphi_{1}^{\prime}\left(
t_{c}\right)  \right]  +t_{c}\varphi_{1}^{\prime\prime}\left(  t_{c}\right)
}. \label{300}%
\end{equation}
Here we used that $\Theta\left(  0\right)  =t_{c}=\varphi_{1}\left(
t_{c}\right)  $ and $-\Theta^{\prime}\left(  0\right)  =t_{c}/\left[
1-\varphi_{1}^{\prime}\left(  t_{c}\right)  \right]  $, which follows from Eq.
(\ref{220}).

\section{Spectral densities and propagators for various networks}

\label{allqm}

Here we indicate four distinct kinds of uncorrelated random networks with
qualitatively different asymptotic behaviors of $T\left(  0,x\right)  \equiv
T_{0}\left(  x\right)  $ at $x\rightarrow\infty$, where $T_{0}\left(
x\right)  $ is the solution of Eq.~(\ref{140}) at $s=0$. At $x=0$, we always
have $T_{0}\left(  0\right)  =1$. At $x\rightarrow\infty$ we have
$T_{0}\left(  \infty\right)  =\lim_{x\rightarrow\infty}\lim_{s\rightarrow
0}T\left(  s,x\right)  =\Theta\left(  0\right)  $. These four types of networks differ from
each other mainly by a value of the minimum vertex degree.

\begin{enumerate}
\item If the minimum vertex degree $q_{m}\geq3$, then identically $\Theta\left(  x\right)
=0$, and $T_{0}\left(  x\right)  $ exponentially decays to
$T_{0}\left(  \infty\right)  =0$ (see Sec.~\ref{qm3}).

\item If $q_{m}=2$, then identically $\Theta\left(  x\right)  =0$, and
$T_{0}\left(  x\right)  $ decays to $T_{0}\left(  \infty\right)  =0$, but
slower than any exponent (see Sec.~\ref{qm2}).

\item If $q_{m}=1$, then there are two possibilities (see
Sec.~\ref{qm1}):

\begin{enumerate}
\item If $z_{1}=\varphi_{1}^{\prime}\left(  1\right)  >1$, then $0<\Theta
\left(  0\right)  =t_{c}=\varphi_{1}\left(  t_{c}\right)<1$, $T_{0}\left(  x\right)  \rightarrow t_{c}$ as
$x\rightarrow+\infty$. In this case the graph has a giant connected component
and a number of finite ones.

\item If $z_{1}=\varphi_{1}^{\prime}\left(  1\right)  <1$, then $\Theta\left(
0\right)  =1$. $T_{0}\left(  x\right)  =1$ as $x>0$. In this case the graph
consists of only finite connected components.
\end{enumerate}
\end{enumerate}

Let us assume $q_{m}>1$ and consider in Eq. (\ref{140}) the case of small
positive $s$ and large $x$. According to definition (\ref{130}), $T\left(
s,x\right)  $ is actually a Laplace transform of the probability distribution of
the non-negative random variable $\tau$. Hence it cannot decay at
$x\rightarrow\infty$ faster than 
exponentially. In Appendix \ref{A} we show 
that
\begin{equation}
T\left(  s,x\right)  \rightarrow A\exp\left[  -\tau_{m}\left(  s\right)
x-\vartheta\left(  s,x\right)  \right]  . \label{310}%
\end{equation}
Here $A$ is simply a constant, and $\vartheta$ is some correction term in the
exponential. The coefficient $\tau_{m}$ at the main, linear in $x$, term in
the exponential turns out to be the same as for regular Bethe lattice. It is
defined by the relation:%
\begin{equation}
\frac{\tau_{m}}{1-\tau_{m}}=s+\left(  q_{m}-1\right)  \tau_{m}. \label{320}%
\end{equation}
This equation has two real solutions as $s>s_{c}$, where
\begin{equation}
s_{c}=-\lambda_{c}=-q_{m}+2\sqrt{q_{m}-1}\leq0. \label{325}%
\end{equation}
The physical branch of $\tau_{m}\left(  s\right)  $ is the branch, positive at
$s>0$. The other term in the exponent in Eq. (\ref{310}) is a sublinear
function of $x$. Namely,
\begin{equation}
\vartheta\left(  x\right)  =Bx^{\alpha},\ \ \alpha=\frac{\ln\left(
q_{m}-1\right)  }{2\ln\left[  1/\left(  1-\tau_{m}\right)  \right]  }%
=\frac{\ln\left[  1/\left(  1-\tau_{c}\right)  \right]  }{\ln\left[  1/\left(
1-\tau_{m}\right)  \right]  }, \label{330}%
\end{equation}
where $B$ is some constant. Here we introduced $\tau_{c}=\tau_{m}\left(
s_{c}\right)  =1-1/\sqrt{q_{m}-1}$.

As $s$ is close to $s_{c}$, $\alpha$ is close to $1$, and $\vartheta\left(
s,x\right)  $ becomes comparable with the main term. It is this region that
determines physically interesting results. The behavior of $\vartheta\left(
s,x\right)  $ at large $x$ and $s$ close to $s_{c}$ determines the behavior of
the spectral density $\rho\left(  \lambda\right)  $ near its edge $\lambda
_{c}=-s_{c}$ and the behavior of the autocorrelator $\bar{p}_{0}\left(
t\right)  $ at large $t$. It turns out (see Appendix \ref{A}), that the
analytic continuation of $\vartheta\left(  s,x\right)  $ on negative
$s<-\lambda_{c}$, $\vartheta\left(  -\lambda,x\right)  $, as a function of $x$
is singular at some $x_{s}\sim1/\left(  \lambda-\lambda_{c}\right)  $.
Therefore the upper limit of integration in Eq. (\ref{150}) is in the upper
half-plane of $x$, $\operatorname{Im}x>0$ for $\operatorname{Im}\lambda>0$ and
vice versa. Then from Eq.~(\ref{90}) we obtain
\begin{equation}
\rho\left(  \lambda\right)  =\int_{-i\infty}^{+i\infty}\frac{dx}{2\pi
i}e^{\lambda x}\varphi\left[  T\left(  -\lambda,x\right)  \right]  .
\label{340}%
\end{equation}
Note that if $q_{m}=2$, then $s_{c}=\tau_{c}=0$. We consider this case separately
in Sec.~\ref{qm2}.

\subsection{Minimum degree $q_{m}>2$}

\label{qm3}

Let us set $T\left(  s,x\right)  =A\exp\left[  -\tau_{c}x-\vartheta\left(
s,x\right)  \right]  $. Note the difference of the definition of $\vartheta$
with that in Eq. (\ref{310}): here we have the term $-\tau_{c}x$ instead of
$-\tau_{m}\left(  s\right)  x$ in the exponent. Therefore $\vartheta^{\prime
}\left(  s,x=+\infty\right)  =\tau_{m}\left(  s\right)  -\tau_{c}\ll1$ now is not
equal to $0$. In Appendix~\ref{A3} we obtain the following expression,
valid when $x\gg1$ and $\left\vert s\right\vert \ll1$:%
\begin{equation}
\vartheta\left(  s,x\right)  =\frac{x\sqrt{s-s_{c}}}{\left(  q_{m}-1\right)
^{3/4}}\coth\left[  \frac{\sqrt{s-s_{c}}\ln\left(  Cy\right)  }{\left(
q_{m}-1\right)  ^{1/4}\ln\left(  q_{m}-1\right)  }\right]  , \label{350}%
\end{equation}
where $C\sim1$ is some number. Replace $s$ by $-\lambda<-\lambda_{c}$,
$\lambda_{c}=-s_{c}=2\sqrt{q_{m}-1}-q_{m}$. Then we have
\begin{equation}
\vartheta\left(  -\lambda,x\right)  =\frac{x\sqrt{\lambda-\lambda_{c}}%
}{\left(  q_{m}-1\right)  ^{3/4}}\cot\left[  \frac{\sqrt{\lambda-\lambda_{c}%
}\ln\left(  Cy\right)  }{\left(  q_{m}-1\right)  ^{1/4}\ln\left(
q_{m}-1\right)  }\right]  . \label{360}%
\end{equation}
This function has a singularity when the argument of $\cot$ equal to $\pi$,
i.e., at $x=x_{0}$, where
\begin{equation}
x_{0}=C^{-1}\exp\left[  \frac{\pi\left(  q_{m}-1\right)  ^{1/4}\ln\left(
q_{m}-1\right)  }{\sqrt{\lambda-\lambda_{c}}}\right]  =C^{-1}\left(
q_{m}-1\right)  ^{\pi\left(  q_{m}-1\right)  ^{1/4}/\sqrt{\lambda-\lambda_{c}%
}}. \label{370}%
\end{equation}
When $x$ is close to $x_{0}$, one can replace $\cot z\rightarrow-1/(\pi-z)$ in Eq.~(\ref{360}).

Since $T\left(  -\lambda,x\right)  $ is small at large $0<x\lesssim\pi$, one
can replace $\varphi\left[  T\left(  -\lambda,x\right)  \right]  $ by its
leading term $\Pi\left(  q_{m}\right)  \varphi^{q_{m}}T\left(  -\lambda
,x\right)  $. Then, changing in Eq. (\ref{340}) the integration variable,
$x=x_{0}y$, and taking into account Eq. (\ref{310}), we obtain up to a
factor $\sim1$:
\begin{equation}
\rho\left(  \lambda\right)  \sim x_{0}\int_{C}\frac{dy}{2\pi i}\exp\left[
-x_{0}y\left(  b+\frac{a}{\ln y}\right)  \right]  ,\ \ a=\frac{q_{m}\ln\left(
q_{m}-1\right)  }{\sqrt{q_{m}-1}},\ \ b=\frac{q_{m}-2}{\sqrt{q_{m}-1}}.
\label{380}%
\end{equation}
Finally, calculating this integral in the saddle point approximation, we
obtain the density of eigenvalues of the Laplacian spectrum near its endpoint
$\lambda_{c}$:%
\begin{equation}
\rho\left(  \lambda\right)  \sim\exp\left[  \frac{\beta}{2\sqrt{\lambda
-\lambda_{c}}}-d\exp\left(  \frac{\beta}{\sqrt{\lambda-\lambda_{c}}}\right)
\right]  ,\ \beta=\pi\left(  q_{m}-1\right)  ^{1/4}\ln\left(  q_{m}-1\right)
, \label{390}%
\end{equation}
where $d$ is some constant. Substituting Eq. (\ref{390}) into the expression
for the autocorrelator (\ref{100}) and using the saddle point approximation to
calculate the integral, we get:%
\begin{equation}
\bar{p}_{0}\left(  t\right)  \sim\exp\left[  -\lambda_{c}t-\frac{\beta^{2}%
t}{\ln^{2}\left(  dt\right)  }\right]  . \label{400}%
\end{equation}

Recall the notation $T_{0}\left(  x\right)  =T\left(  0,x\right)  $. Since
$\tau_{m}\left(  0\right)  =\left(  q_{m}-2\right)  /\left(  q_{m}-1\right)
>0$, we have $\varphi_{1}^{\prime}\left[  T_{0}\left(  x\right)  \right]
\sim\exp\left[  -\left(  q_{m}-2\right)  \tau_{m}\left(  0\right)  x\right]  $
at $x\rightarrow+\infty$, and the kernel of the integral equation (\ref{2020})
satisfies the condition (\ref{2025}). It implies that at $s=0$ in the discrete
sequence of characteristic numbers $\mu_{m}\left(  0\right)  \equiv\mu_{m}$, 
there is the minimum one, $\mu_{0}>0$. In Appendix \ref{P} we show that (i)
$\mu_{0}=1$, (ii) this characteristic number is the minimum one, and (iii) the
corresponding normalized eigenfunction is
\begin{equation}
\psi_{0}\left(  0,x\right)  \equiv\psi_{0}\left(  x\right)  =-d_{0}%
T_{0}^{\prime}\left(  x\right)  ,\ d_{0}=\left[\int_{0}^{\infty
}dx~\varphi_{1}^{\prime}\left[  T_{0}\left(  x\right)  \right]  T_{0}%
^{\prime2}\left(  x\right)  \right]  ^{-1/2}. \label{410}%
\end{equation}
Here $d_{0}$ ensures proper normalization (\ref{2030}), and the minus sign
stands simply for convenience ensuring $\psi_{0}\left(  x\right)  \geq0$.

When $s>s_{c}$, in particular, near $s=0>s_{c}$, the kernel in the integral
equation (\ref{2020}) is well-behaved, and all$\ \mu_{m}\left(  s\right)  $
are analytic functions of $s$. We can leave in Eq. (\ref{2060}) only the
leading term with the minimum $\mu_{m}$. Then $\bar{P}_{l}\left(  s\right)
\approx A_{0}\left(  s\right)  B_{0}\left(  s\right)  \mu_{0}^{1-l}\left(
s\right)  $ for large distances $l$ from the initial vertex. So at large time
$t$ and large distance $l$, the propagator $\bar{p}_{l}\left(  t\right)  $ is
approximately
\begin{equation}
\bar{p}_{l}\left(  t\right)  =\int_{-i\infty}^{+i\infty}\frac{ds}{2\pi
i}e^{st}\bar{P}_{l}\left(  s\right)  \approx\int_{-i\infty}^{+i\infty}%
\frac{ds}{2\pi i}e^{st}A_{0}\left(  s\right)  B_{0}\left(  s\right)  \mu
_{0}^{1-l}\left(  s\right)  . \label{420}%
\end{equation}
If the expression under the integral is analytic in $s$ along the integration
contour, the main contribution to the asymptotic of the integral gives the
vicinity of the saddle point, where $st-l\ln\mu_{0}\left(  s\right)  $ is
maximal. The saddle point position $s_{c}$ is the solution of the equation
$t=l\mu_{0}^{\prime}\left(  s\right)  /\mu_{0}\left(  s\right)  $. As a
result, we have
\begin{equation}
\bar{p}_{l}\left(  t\right)  \approx\frac{1}{\sqrt{2\pi\beta\left(
s_{c}\right)  l}}A_{0}\left(  s_{c}\right)  B_{0}\left(  s_{c}\right)  \mu
_{0}\left(  s_{c}\right)  \exp\left[  s_{c}t-l\ln\mu_{0}\left(  s_{c}\right)
\right]  , \label{425}%
\end{equation}
where $\beta\left(  s_{c}\right)  =\left.  \left[  \ln\mu_{0}\left(  s\right)
\right]  ^{\prime\prime}\right\vert _{s=s_{c}}$. At a given $t\gg1$, this
expression has a maximum as a function of $l=l_{m}\left(  t\right)  $ at the
point where
\[
\frac{\partial}{\partial l}\left[  s_{c}t-l\ln\mu_{0}\left(  s_{c}\right)
\right]  =\ln\mu_{0}\left(  s_{c}\right)  =0,
\]
i.e., where $\mu_{0}\left[  s_{c}\left(  l_{m},t\right)  \right]  =1$. Here
$s_{c}\left(  l,t\right)  $ is defined from the saddle point condition. Since
$\mu_{0}\left(  0\right)  =1$, the propagator $\bar{p}%
_{l}\left(  t\right)  $ is maximal at $l=l_{m}$: $s_{c}\left(  l_{m},t\right)
=0$. The behavior of $\mu_{0}\left(  s\right)  $ at small values of
$\left\vert s\right\vert $ determines the shape of the propagator near its
maximum point. Since $\mu_{0}\left(  s\right)  $ is an analytic function near
$s=0$, and $\mu_{0}\left(  0\right)  =1$, one can write $\ln\mu_{0}\left(
s\right)  =\alpha s-\beta s^{2}/2+\cdots$ and replace $A_{0}\left(  s\right)
$ and $B_{0}\left(  s\right)  $ by $A_{0}\left(  0\right)  $ and $B_{0}\left(
0\right)  $. Then the expression (\ref{425}) is reduced to a Gaussian integral,
and we have
\begin{equation}
\bar{p}_{l}\left(  t\right)  \approx\frac{A_{0}\left(  0\right)  B_{0}\left(
0\right)  }{\sqrt{2\pi\beta l}}\exp\left[  -\frac{\left(  t-\alpha l\right)
^{2}}{2\beta l}\right]  . \label{430}%
\end{equation}
On the left-hand side of the normalization condition (\ref{2080}), only the
term with $m=0$ has a simple pole singularity at $s=0$. Then we have
$\lim_{s\rightarrow0}A_{0}\left(  s\right)  B_{0}\left(  s\right)  /\left[
\mu_{0}\left(  s\right)  -1\right]  =A_{0}\left(  0\right)  B_{0}\left(
0\right)  /\alpha=1$. We substitute the expressions for $A_{0}\left(
0\right)  $ and $B_{0}\left(  0\right)  $ from Eqs. (\ref{2050}) and
(\ref{2070}), where the function $\psi_{0}\left(  x\right)  $ is expressed in
terms of $T_{0}\left(  x\right)  $ by using Eq. (\ref{410}). This leads to
\begin{equation}
\alpha\equiv v^{-1}=A_{0}\left(  0\right)  B_{0}\left(  0\right)  =\frac
{\int_{0}^{\infty}dx~T_{0}^{\prime}\left(  x\right)  \varphi_{1}^{\prime
}\left[  T_{0}\left(  x\right)  \right]  \left(  1+\frac{d}{dx}\right)
T_{0}\left(  x\right)  \int_{0}^{\infty}dx~\varphi^{\prime}\left[
T_{0}\left(  x\right)  \right]  T_{0}^{\prime}\left(  x\right)  }{\int
_{0}^{\infty}dx~\varphi_{1}^{\prime}\left[  T_{0}\left(  x\right)  \right]
T_{0}^{\prime2}\left(  x\right)  }. \label{440}%
\end{equation}
The parameter $\beta\sim1$ must be positive to ensure the convergence in the
summation over $l$. Equation~(\ref{430}), as one can see from its derivation,
is valid if the saddle point position $|s_{c}| =|(t-l/v)/(\beta l)| \ll l$. So we may replace $\beta l$ in
Eq.~(\ref{430}) with its value at $l=l_{m}$, $\beta vt$, and, finally,
\begin{equation}
\bar{p}_{l}\left(  t\right)  \approx\frac{1}{\sqrt{2\pi Dt}}\exp\left[
-\frac{\left(  l-vt\right)  ^{2}}{2Dt}\right]  , \label{450}%
\end{equation}
where $D=\beta v^{3}$. Despite our network is random, a signal spreads over
the network as a Gaussian packet, moving with the constant velocity $v$ from
an initial vertex, and with the dispersion $\overline{\left(  l-l_{m}\right)
^{2}}$, which grows linearly with time. This is the same kind of evolution as
on a regular Bethe lattice.

Equation (\ref{450}) is valid when one can neglect terms of the order of
$s^{3}$ and higher in the expansion of $\ln\mu_{0}\left(  s\right)  $ in the
powers of $s$, i.e., $l\left\vert s\right\vert _{c}^{3}\sim t\left\vert
s\right\vert _{c}^{3}\ll1$. Since $s_{c}=\left(  l/v-t\right)  /l\sim\left(
l-vt\right)  /t$, this condition is reduced to $\left\vert l-vt\right\vert \ll
t^{2/3}$. The width of the packet is $\sim t^{1/2}\ll t^{2/3}$, and so
expression~(\ref{450}) is relevant.

\subsection{Minimum degree $q_{m}=2$}

\label{qm2}

If $q_{m}=2$, then $s_{c}=0$ as one can see from Eq. (\ref{325}). That is,
$T\left(  s,x\right)  $ becomes nonanalytic at $s<0$. Besides, $\tau_{c}%
=\tau_{m}\left(  s_{c}\right)  =0$, so that the decay of $T_{0}\left(
x\right)  \equiv T(0,x)$ is nonexponential in contrast to $q_{m}>2$. Setting
$T\left(  s,x\right)  =\exp\left[  -\vartheta\left(  s,x\right)  \right]  $,
we obtain the following expression for small $s$ and large $x>0$ (see Appendix
\ref{A2}):%
\begin{equation}
\vartheta\left(  s,x\right)  \approx\frac{1}{\sqrt{s}}\left[  \sqrt{sx\left(
a/\pi+sx\right)  }+\frac{a}{\pi}\operatorname{arcsinh}\sqrt{\frac{\pi sx}{a}%
}\right]  +\frac{1}{4}\ln\left(  s+\frac{a}{\pi x}\right)  +C,\ \ \ a=\pi
\ln\left[  \frac{\bar{q}}{2\Pi\left(  2\right)  }\right]  >0, \label{460}%
\end{equation}
where $C\sim1$ is some constant. In the following we omit numerical
constants as inessential. When analytically continued to $s=-\lambda<0$,
$\vartheta\left(  -\lambda,x\right)  $ as a function of $x$ acquires a
singularity at $x=x_{c}=a/\pi\lambda$. The density of Laplacian eigenvalues,
$\rho\left(  \lambda\right)  ,$ can be obtained from Eq. (\ref{340}). The main
contribution to the integral in Eq.~(\ref{340}) arises from the close vicinity
of the singularity point. In Eq.~(\ref{340}), we expand $\vartheta$ near
$x_{c}$ in the integral and change the integration variable from $x$ to
$\zeta=\lambda\left(  x_{c}-x\right)  $. This results in%
\[
\rho\left(  \lambda\right)  \sim\frac{1}{\lambda^{3/2}}\exp\left(  -\frac
{a}{\sqrt{\lambda}}\right)  \int_{C^{\prime}}\frac{d\zeta}{2\pi i}\zeta
^{-1/2}\exp\left(  -\zeta+\frac{4\zeta^{3/2}}{3\sqrt{\lambda}}\right)  .
\]
Here the integral term is $\sim\lambda^{1/6}$, and the asymptotics at
$0<\lambda\ll1$ is:%
\begin{equation}
\rho\left(  \lambda\right)  \sim\frac{1}{\lambda^{4/3}}\exp\left(  -\frac
{a}{\sqrt{\lambda}}\right)  . \label{470}%
\end{equation}
We substitute this expression into Eq.~(\ref{100}), and by using the saddle
point approximation, arrive at the following long $t$ asymptotics for the
autocorrelator:%
\begin{equation}
\bar{p}_{0}\left(  t\right)  \sim t^{1/18}\exp\left[  -3\left(  \frac{a}%
{2}\right)  ^{2/3}t^{1/3}\right]  . \label{480}%
\end{equation}

Let us now consider the propagator $\bar{p}_{l}\left(  t\right)  $ at $l\gg1$,
$t\gg1$. This asymptotics is also defined by Eq. (\ref{420}). As for $q_{m}%
>2$, the main contribution to the integral is from the region of small
$\left\vert s\right\vert $. The difference is that here $A_{0}\left(
s\right)  $, $B_{0}\left(  s\right)  $ and $\mu_{0}\left(  s\right)  $ all
have a singularity at $s=0$. Namely, $s=0$ is a branching point, giving a cut
along the line $(0,\infty)$ in the complex plane of the variable $s$. We will show, however,
that this singularity is very weak and does not contribute essentially to the 
propagator, except of relatively small distances $l$.

Indeed, the small $s$, large $x$ asymptotics of the eigenfunction $\psi
_{0}\left(  s,x\right)  $, corresponding to the largest characteristic number
$\mu_{0}\left(  s\right)  =1+o\left(  s\right)  $, is (see Appendix \ref{A2}):%
\begin{equation}
\psi_{0}\left(  s,x\right)  \approx x^{-1/2}T\left(  s,x\right)  \sim
x^{-1/2}\left(  s+\frac{a}{\pi x}\right)  ^{-1/4}\exp\left\{  -\frac{1}%
{\sqrt{s}}\left[  \sqrt{sx\left(  a/\pi+sx\right)  }+\frac{a}{\pi
}\operatorname{arcsinh}\sqrt{\pi sx/a}\right]  \right\}  . \label{482}%
\end{equation}
Then, comparing the leading terms in Eqs. (\ref{2050}) and (\ref{2070}) with
that in Eq. (\ref{150}), we conclude that the asymptotics of
$\operatorname{Im}A_{0}\left(  -\lambda\right)  $ and of $\operatorname{Im}%
B_{0}\left(  -\lambda\right)  $ on $\lambda$ are nearly the same as that of
$\rho\left(  \lambda\right)  \sim\operatorname{Im}\bar{P}_{0}\left(
-\lambda\right)  $. The difference is in powers of $\lambda$ in the
pre-exponential factors. In the leading order,
\begin{equation}
\operatorname{Im}\left[  A_{0}\left(  -\lambda\right)  \right]  \sim
\operatorname{Im}\left[  B_{0}\left(  -\lambda\right)  \right]  \sim
\int_{-i\infty}^{+i\infty}\frac{dx}{2\pi i}e^{\lambda x}T\left(
-\lambda,x\right)  \psi_{0}\left(  -\lambda,x\right)  ~\sim\lambda^{-5/6}%
\exp\left(  -\frac{a}{\sqrt{\lambda}}\right)  . \label{484}%
\end{equation}
The rate of singularity of $\mu_{0}\left(  s\right)  $, if measured as a jump
of a function across the cut near its branching point, is even smaller than in Eq.~(\ref{484}) for
small $\lambda=-s>0$. Let us take the eigenfunction equation (\ref{2020}) at
$m=0$, setting $x=0$. Then we have
\[
\mu_{0}^{-1}\left(  s\right)  \psi_{0}\left(  s,0\right)  =\int_{0}^{\infty
}dy~e^{-\left(  1+s\right)  y}\varphi_{1}^{\prime}\left[  T\left(  s,y\right)
\right]  \psi_{0}\left(  s,y\right)  .
\]
Then, setting $s=-\lambda<0$, and properly deforming integration contour, we obtain in the leading order:%
\[
\operatorname{Im}\psi\left(  -\lambda,0\right)  -\psi\left(  0,0\right)
\operatorname{Im}\mu_{0}\left(  -\lambda\right)  \sim\int_{C}\frac{dx}%
{\sqrt{x}}~\exp\left[  -\left(  1-\lambda\right)  x-\frac{1}{\sqrt{\lambda}%
}\vartheta\left(  -\lambda,x\right)  \right]  ,
\]
where the function $\vartheta$ is given by Eq. (\ref{460}). As a function of
$x$ this integral has a singularity at $x=a/\pi\lambda$. In comparison with the integral
for $\rho\left(  \lambda\right)  $, the above integral has an additional term
$-x$ in the exponent, which turns into $-a/\pi\lambda$ at the singularity
point. Therefore, we estimate the singularity of $\mu_{0}$ near $s=0$ as
$\operatorname{Im}\mu_{0}\left(  -\lambda\right)  \sim\exp\left(
-a/\pi\lambda\right)  $.

Since all multipliers in Eq. (\ref{420}) have sufficiently weak singularities,
we replace $A_{0}\left(  s\right)  $ and $B_{0}\left(  s\right)  $ with their
values at $s=0$ and neglect the singular part of $\mu_{0}\left(  s\right)  $,
leaving only the regular part of the expansion: $\ln\mu_{0}\left(  s\right)
=s/v+\beta s^{2}/2+\cdots$. As a result, we arrive at the same Gaussian
expression for the propagator, Eq. (\ref{450}).

If we, however, fix the distance $l\gg1$ and increase the time $t$, the saddle
point $s_{c}<0$ in the integral (\ref{420}) moves farther in the direction of
negative $s$, and at large enough $t$ the contribution of the singularity
becomes essential. Deforming contour of integration, we rewrite Eq.
(\ref{420}) in the following form:
\begin{equation}
\bar{p}_{l}\left(  t\right)  \approx\frac{1}{\pi}\int_{0}^{\infty}%
d\lambda~e^{-\lambda t}\operatorname{Im}\left[  A_{0}\left(  -\lambda\right)
B_{0}\left(  -\lambda\right)  \mu_{0}^{1-l}\left(  -\lambda\right)  \right]  .
\label{4840}%
\end{equation}
$\operatorname{Im}\mu_{0}\sim\exp\left(  -a/\pi\lambda\right)  $ is
small compared to $\operatorname{Im}A_{0}\left(  -\lambda\right)
\sim \operatorname{Im}B_{0}\left(  -\lambda\right)  \sim\exp(-a/\sqrt
{\lambda})$. So we neglect the singularity of $\mu_{0}$ and set $\ln\mu_{0}\left(  -\lambda\right)  =-\lambda/v$. Thus we arrive at
\begin{equation}
\bar{p}_{l}\left(  t\right)  \sim\int_{0}^{\infty}\frac{d\lambda}%
{\lambda^{5/6}}\exp\left[  -\lambda\left(  t-\frac{l}{v}\right)  -\frac
{a}{\sqrt{\lambda}}\right]  . \label{485}%
\end{equation}
Calculating the integral in the saddle point approximation we obtain
\begin{equation}
\bar{p}_{l}\left(  t\right)  \sim\left(  t-\frac{l}{v}\right)  ^{-5/18}%
\exp\left[  -3\left(  \frac{a}{2}\right)  ^{2/3}\left(  t-\frac{l}{v}\right)
^{1/3}\right]  . \label{486}%
\end{equation}
Expanding $\ln\mu_{0}\left(  s\right)  $, we neglected terms of the order of
$s^{2}$ and higher. This is justified if the saddle point position in the
integral (\ref{485}), $\lambda_{s}\sim\left(  t-l/v\right)  ^{-2/3}$, obeys
the condition $l\lambda_{s}^{2}\ll\lambda_{s}^{-1/2}$ which is equivalent to
$t-l/v\gg t^{3/5}$. Otherwise, $\bar{p}_{l}\left(  t\right)  $ is given by Eq.
(\ref{450}), which means that the probability for the signal to return is
small. This form of the packet tail is due to the possibility that either
initial vertex $0$ or the final one in the $l$-th shell may occur in a 
chain fragment in the graph.

\subsection{Minimum degree $q_{m}=1$}

\label{qm1}

When there is a finite fraction of \textquotedblleft dead
ends\textquotedblright, i.e., vertices of degree $1$, the network contains
finite-size connected components. They lead to the $\delta$-functional peak in
the Laplace spectrum and so to nonzero limits of the averaged propagators at
$t\rightarrow\infty$. If $\varphi_{1}^{\prime}\left(  1\right)  >1$ (Appendix
\ref{Z}), then besides the finite connected components, there is a giant
connected one whose size scales as the network size. Here we show that
the contribution of this giant connected component to the observable
quantities is qualitatively the same as in networks with $q_{m}=2$.

\begin{figure}[t]
\begin{center}
\scalebox{0.63}{\includegraphics[angle=0]{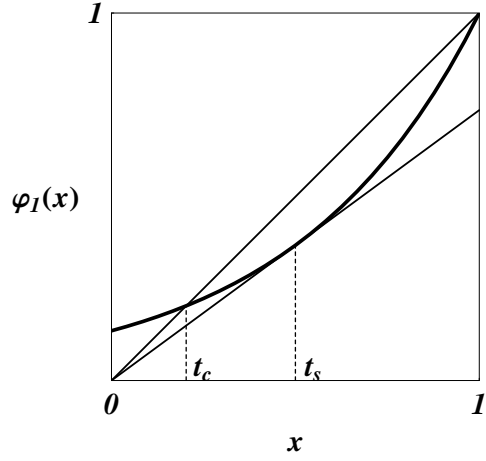}}
\end{center}
\par
\caption{ Graphical solution of the equations for $t_{c}$, $t_{c}=\varphi
_{1}\left(  t_{c}\right)  $ and for $t_{s}>t_{c}$: $\varphi_{1}^{\prime
}\left(  t_{s}\right)  =\varphi_{1}\left(  t_{s}\right)  /t_{s}$.}%
\label{phi1}%
\end{figure}

If $q_{m}=1$, Eq.~(\ref{140}) still has the nontrivial solution
$T_{0}\left(  x\right)  \equiv T\left(  0,x\right)  $. $T_{0}\left(  0\right)
=1$ as for any other $q_m$, but
$T_{0}\left(  +\infty\right)  =\lim_{x\rightarrow0}\lim_{s\rightarrow
0}T\left(  s,x/s\right)  =\Theta\left(  0\right)  =t_{c}>0$ (see Appendix
\ref{Z}). At small $s>0$ and large $x>0$, the function $T\left(  s,x\right)  $
is close to $\Theta\left(  sx\right)  $, and so we search for $T\left(
s,x\right)  $ in the following form:
\begin{equation}
T\left(  s,x\right)  =\Theta\left(  sx\right)  +e^{-\vartheta\left(
s,x\right)  }, \label{490}%
\end{equation}
where the last term is assumed to be small. The asymptotic solution for
$\vartheta$ is (Appendix \ref{A1})
\begin{equation}
\vartheta\left(  s,x\right)  =\frac{1}{\sqrt{s}}g\left(  sx\right)  +\frac
{1}{4}\ln sg^{\prime^{2}}\left(  sx\right)  +C, 
\ \ g\left(  z\right)  =\int
_{0}^{z}dy\sqrt{1-\frac{\ln\varphi_{1}^{\prime}\left[  \Theta\left(  y\right)
\right]  }{y}}, \label{500}%
\end{equation}
where $C\sim1$ is some constant. Continuing this result to $s=-\lambda>0$, we
take into account that $g\left(  z\right)  $ has a singularity at $z=z_{s}<0$,
where $z_{s}$ satisfies the equation $1-\ln\varphi_{1}^{\prime}[\Theta(z_{s})]/z_{s}=0$. The
equation for $z_{s}$, $\varphi_{1}^{\prime}\left[  \Theta\left(  z_{s}\right)
\right]  =e^{z_{s}}$, becomes more comprehensive with a new variable
$t_{s}=\Theta\left(  e^{-z_{s}}\right)  $. Using the implicit definition
(\ref{220}) of $\Theta\left(  z\right)  $, we arrive at the equation for
$t_{s}$:%
\begin{equation}
\varphi_{1}^{\prime}\left(  t_{s}\right)  =\frac{\varphi_{1}\left(
t_{s}\right)  }{t_{s}}. \label{510}%
\end{equation}
This equation is shown graphically in Fig. \ref{phi1}, together with the
equation for $t_{c}=\varphi_{1}\left(  t_{c}\right)  <t_{s}$, $\Theta\left(
t_{c}\right)  =0$.

An expression for the spectral density $\rho\left(  \lambda\right)  $
may be obtained by calculating the integral in Eq. (\ref{340}), where for
small $\lambda$ and large $x$, we approximately set:%
\begin{equation}
\varphi\left[  T\left(  -\lambda,x\right)  \right]  \approx\varphi\left[
\Theta\left(  -\lambda x\right)  \right]  +\varphi^{\prime}\left[
\Theta\left(  -\lambda x\right)  \right]  e^{-\vartheta\left(  -\lambda
,x\right)  }.\label{520}%
\end{equation}
While the first term results in the $\delta$-functional peak, the integration
of the last one gives the asymptotics of $\rho\left(  \lambda\right)  $ at
small positive $\lambda$. The main contribution into the integral gives the
close vicinity of the positive singularity point $x_{s}=-z_{s}/\lambda>0$.
Near this point the first multiplier in the last term in Eq. (\ref{520}) can be
replaced by a constant $\varphi^{\prime}\left[  \Theta\left(  z_{s}\right)
\right]  =\varphi^{\prime}\left(  t_{s}\right)  $, and the function $g$ in the
expression for $\vartheta$ (\ref{500}) can be replaced by its expansion
$g\left(  z_{s}+\eta\right)  =ia+ih\eta^{5/4}$, $h\sim1$, $\left\vert
\eta\right\vert \ll1$ with%
\begin{equation}
a=\int_{t_{c}}^{t_{s}}dx\frac{\varphi_{1}\left(  x\right)  -x\varphi
_{1}^{\prime}\left(  x\right)  }{x\varphi_{1}\left(  x\right)  }\sqrt
{\frac{\ln\varphi_{1}^{\prime}\left(  x\right)  }{\ln\left[  \varphi
_{1}\left(  x\right)  /x\right]  }-1}\label{530}%
\end{equation}
(see Appendix \ref{A1}). Then we arrive at the following asymptotic result for
$\rho\left(  \lambda\right)  $:%
\begin{equation}
\rho\left(  \lambda\right)  -p_{0}^{\left(  \mathrm{eq}\right)  }\delta\left(
\lambda\right)  \sim\lambda^{-9/10}\exp\left(  -\frac{a}{\sqrt{\lambda}%
}\right)  .\label{540}%
\end{equation}
As it follows from Eqs. (\ref{540}) and (\ref{100}), the autocorrelator
$\bar{p}_{0}\left(  t\right)  $ decays to its equilibrium value as
\begin{equation}
\bar{p}_{0}\left(  t\right)  -p_{0}^{\left(  \mathrm{eq}\right)  }\sim\int
_{0}^{\infty}\frac{d\lambda}{\lambda^{9/10}}\exp\left(  -\lambda t-\frac
{a}{\sqrt{\lambda}}\right)  \sim t^{-7/30}\exp\left[  -3\left(  \frac{a}%
{2}\right)  ^{2/3}t^{1/3}\right]  .\label{550}%
\end{equation}

One can calculate the propagator $\bar{p}_{l}\left(  t\right)  $ at large $l$
using Eq. (\ref{420}). As compared with Secs.~\ref{qm3} and \ref{qm2}, the
kernel of Eq.~(\ref{2020}) is not any more bounded at $s=0$ because
the integral in Eq. (\ref{2025}) becomes divergent. Due to this fact, the
spectrum of Eq.~(\ref{2020}) contains continuous part. Let us find
eigenvalues $\mu_{\xi}$ and eigenfunctions $\psi_{\xi}\left(  x\right)  $ in
the continuous spectrum. (The notation $\psi_{m}\left(  x\right)  \equiv
\psi_{m}\left(  s=0,x\right)  $ we leave for the discrete part of the
spectrum.) We saw in Sec.~\ref{finite} that when $s\rightarrow0$, the
recursion relation (\ref{180}) can be transformed to Eq. (\ref{260}), assuming
that $U_{l}\left(  s,x\right)  \approx u_{l}\left(  sx\right)  $ at small $s$.
An equation for the eigenfunctions is
\[
\psi_{\xi}\left(  x\right)  =\mu_{\xi}e^{-x}\varphi_{1}^{\prime}\left[
\Theta\left(  x\right)  \right]  \psi_{\xi}\left(  x\right)  ,
\]
which has the solutions $\psi_{\xi}\left(  x\right)  =\delta\left(
x-\xi\right)  $ corresponding to the eigenvalues $\mu_{\xi}=e^{\xi}%
/\varphi_{1}^{\prime}\left[  \Theta\left(  x\right)  \right]  $. It is the
continuous part of the spectrum that after proper modification of the
relations (\ref{2030}-\ref{2070}), gives the stationary part of propagator
(\ref{290}). Suppose that there is a giant connected component in the network.
Then along with the continuous part of the spectrum, whose minimum
characteristic number is $\mu_{\xi=0}=1/\varphi_{1}^{\prime}\left(
t_{c}\right)  >1=\mu_{m=0}$, there is a discrete spectrum with the minimum
characteristic number $\mu_{0}\left(  s=0\right)  =1$ corresponding to the
eigenfunction $\psi_{0}\left(  x\right)  =-T_{0}^{\prime}\left(  x\right)  $
(see Appendix \ref{P}).

In the same way as for $q_{m}=2$ (see Appendix \ref{A1}), one can show that
the asymptotics at small $s$ and large positive $x$ of the eigenfunction
$\psi_{0}\left(  s,x\right)  $, corresponding to the lowest eigenvalue
$\mu_{0}\left(  s\right)  $, is $\psi_{0}\left(  s,x\right)  \sim x^{-1/2}%
\exp\left[  -\vartheta\left(  s,x\right)  \right]  $, where $\vartheta\left(
s,x\right)  $ is given by Eq. (\ref{500}).

From Eqs. (\ref{2050}) and (\ref{2070}) we obtain
\begin{equation}
\operatorname{Im}\left[  A_{0}\left(  -\lambda\right)  \right]  \sim
\operatorname{Im}\left[  B_{0}\left(  -\lambda\right)  \right]  \sim
\int_{-i\infty}^{+i\infty}\frac{dx}{2\pi i}e^{\lambda x}\psi_{0}\left(
-\lambda,x\right)  \sim\int_{-i\infty}^{+i\infty}\frac{dx}{\sqrt{x}}e^{\lambda
x-\vartheta\left(  -\lambda,x\right)  }\sim\lambda^{-1/2}\exp\left(  -\frac
{a}{\sqrt{\lambda}}\right)  . \label{560}%
\end{equation}
This equation differs from Eq. (\ref{484}), because in the singularity point
$x_{s}=-z_{s}/\lambda\gg1$, $T\left(  -\lambda,x_{s}\right)  \approx
\Theta\left(  -z_{s}\right)  \sim1$. So we omitted $T\left(  -\lambda
,x_{s}\right)  $ in Eq. (\ref{560}), in contrast to the case $q_{m}=2$, where
the function $T\left(  -\lambda,x_{s}\right)  $ has is of the same order of smallness 
as $\psi_{0}\left(  -\lambda,x_{s}\right)  $. Here, as for $q_{m}=2$, the
singularity of $\mu_{0}\left(  s\right)  $ is such that the jump along the cut
$\left(  -\infty,0\right)  $ in the complex planes $s$ behaves as
$\operatorname{Im}\mu_{0}\left(  -\lambda\right)  \sim\exp\left(
-a/\lambda\right)  $.

The derivation of $\bar{p}_{l}\left(  t\right)  $ for $q_{m}=1$ is similar to
that for $q_{m}=2$. We arrive at the same moving Gaussian packet (\ref{450}).
The only difference is that now we must take into account the contribution of
finite clusters (continuous spectrum). The results for $l\gg1$ and $t\gg1$ are
\begin{equation}
\bar{p}_{l}\left(  t\right)  =p_{l}^{\left(  \mathrm{eq}\right)  }+\frac
{1}{\sqrt{2\pi Dt}}\exp\left[  -\frac{\left(  l-vt\right)  ^{2}}{2Dt}\right]
\label{570}%
\end{equation}
for $\left\vert vt-l\right\vert \ll t^{3/5}$, where $v$ is given by Eq.
(\ref{300}), and $D=v^{3}\mu_{0}^{\prime\prime}\left(  0\right)  $. In the
low $l$ tail, $l<vt$, $vt-l\gg t^{3/5}$, the form of the propagator is
modified to
\begin{equation}
\bar{p}_{l}\left(  t\right)  -p_{l}^{\left(  \mathrm{eq}\right)  }\sim\left(
t-\frac{l}{v}\right)  ^{-13/30}\exp\left[  -3\left(  \frac{a}{2}\right)
^{2/3}\left(  t-\frac{l}{v}\right)  ^{1/3}\right]  .\label{580}%
\end{equation}
Thus, again, we have the Gaussian packet, Eq. (\ref{570}), moving within the giant
connected component. This Gaussian is supplied with a small tail at $1 \ll l \ll t$, Eq.
(\ref{580}). The reason for this tail is that initial or final vertices may be
\textquotedblleft dead ends\textquotedblright.

\section{Summary, discussion, and conclusions}

\label{conclusions}

In this article we have presented a theory, which enables us the analytical
calculation of statistical properties of the Laplacian operators of infinite
random networks and random walks on them. We have considered the resolvent of
the Laplacian and the propagator of a random walk. These characteristics are
connected through a Laplace transform, Eqs. (\ref{65}) and (\ref{80}). In
particular, the average values of the diagonal element of the resolvent matrix
give us the spectral density of the Laplacian, Eq. (\ref{90}), and the time
dependence of the autocorrelator. We have also derived equations, which
solution allows us to find the averages of the nondiagonal elements of the
resolvent. After the Laplace transformation, these averages show how the
distance of the signal from its origin changes with time, Eq. (\ref{60}).

Our scheme is based on equations relating the distributions (or other
statistical properties) of random variables. This is an essential advantage
over most of existing approaches, based on equations relating the values of
some random variables for a given network realization. To solve the problems
of the Laplacian spectrum and of random hopping motion, one must make the
following steps.

\begin{itemize}
\item[(i)] Solve the integral equation (\ref{140}) for the function $T\left(
s,x\right)  $ defined by Eqs. (\ref{120}) and (\ref{130}). [In the equivalent
form, it is Eq.~(\ref{T70}).] Technically, it is the most difficult step. We
have only obtained the asymptotics of $T(s,x)$ at $\operatorname{Re}%
x\rightarrow+\infty$. We have found that $T$ as a function of $x$ is an
analytic and exponentially decaying function as $s>s_{c}$, where $s_{c}=-\lambda
_{c}\leq0$ is a parameter which depends only on the minimum vertex degree
$q_{m}$.

\item[(ii)] With $T\left(  s,x\right)  $, one can (a) calculate the average of
the resolvent's diagonal, Eq.~(\ref{150}), then (b) analytically continue the
result from the positive $s$ to $s=-\lambda\pm i0$, $\lambda>0$, and finally
(c) obtain, using Eq.~(\ref{90}), the spectral density of the Laplacian
\cite{remark1}.

\item[(iii)] With the known $\rho(\lambda)$ near the spectrum edge, obtain the
asymptotics of the autocorrelator $\bar{p}_{0}\left(  t\right)  $ at
$t\rightarrow\infty$ by using Eq. (\ref{100}).

\item[(iv)] Find the sequence of functions $U_{l}\left(  s,x\right)  $,
$l\geq1$, definition (\ref{170}), by using the integral recursive relation
(\ref{180}) with the initial condition (\ref{190}). Then obtain the
Laplace-transformed propagator $\bar{P}_{l}\left(  s\right)  $ by calculating
the integral (\ref{200}) \cite{remark2}.

\item[(v)] Calculate the inverse Laplace transform of $\bar{P}_{l}\left(
s\right)  $, that is, the propagator $\bar{p}_{l}\left(  t\right)  $. The
asymptotics of $\bar{p}_{l}\left(  t\right)  $ at large $l$ and large $t$ is
determined by the smallest characteristic number $\mu_{0}\left(  s\right)  $
at small $\left\vert s\right\vert $.
\end{itemize}

The results of these calculations of asymptotics are summarized in Table
\ref{results} and Fig.~\ref{spectra}. If $q_{m}\geq3$, the tail in the density of eigenvalues
decreases extremely rapidly with $1/\left(  \lambda-\lambda_{c}\right)  $, see
Eq. (\ref{390}), and therefore practically cannot be revealed by numerical
methods. Studies based on these methods usually result in a form of
$\rho(\lambda)$ resembling Wigner's semi-circle law (see, e.g.,
Refs.~\cite{clv03,jm}). This is also the case in networks with $q_{m}=1,2$.

When are our analytical results observable? Let us inspect the resulting
expressions for the propagator $\bar{p}_{l}\left(  t\right)  $. Our results
are based on the tree ansatz: $p_{l}\left(  t\right)  $ should have nonzero
values in the small (compared to the whole network) vicinity of the starting
vertex $0$, so that we can treat this region as a tree. At large times $t$ the
signal spreads at the distance $\bar{l}=vt\sim t$, Eq. (\ref{450}). The mean
intervertex distance in the network is $\sim\ln N~$\ \cite{nsw01,dms03}. So,
our results are applicable if $1\ll t\ll\ln N$. In networks with $q_{m}\geq3$
the decay of the autocorrelator is basically exponential with some correction
[see Eq.~(\ref{400})]. This correction can be observed if
\begin{equation}
1/\ln^{2}t\sim1/\ln^{2}\ln N\ll1. \label{590}%
\end{equation}
It seems to be impossible to fulfil this criterion either in real-world
networks or in numerical simulations.

In the networks, containing chain-like segments, i.e., when $q_{m}=1$ or $2$, the
criterion is much less stringent. We require that the value of the
autocorrelator $\bar{p}_{0}\left(  t\right)  $ (\ref{480}) at the
characteristic time $t\sim\ln N$, essentially exceed its equilibrium value
$\bar{p}_{0}\left(  t=\infty\right)  \sim1/N$ for a finite network. So in
these networks, our dependences are observable if
\begin{equation}
t^{1/3}/\ln N\sim\ln^{-2/3}N\ll1, \label{590a}%
\end{equation}
which is much easier to satisfy than 
condition (\ref{590}).

In many applications of the Laplacian spectrum, results, obtained in the
infinite network limit, are of little use. A good example is synchronization
\cite{aj04,adp06}. In this problem the lowest, size-dependent eigenvalue of
the Laplacian plays a key role. Let us briefly discuss the role of this
eigenvalue in application to our problems. The process of a signal spread over
the network consists of two distinct stages. We discussed the first one. In
the second stage, the essence of the process is the relaxation to the
homogeneous distribution, where the probability to find a signal at any vertex
is the same, namely, $1/N$. In this last stage, $\left\vert p_{i0}\left(
t\right)  -1/N\right\vert \ll1/N$. In this situation loops must be taken into
account. Furthermore, in this stage, the knowledge of the Laplacian spectral
density is not sufficient. Rather, one should ask: what is the probability
distribution of $\lambda_{2}$ (the lowest nonzero eigenvalue)?

We showed that in infinite networks with minimum vertex degree $q_{m}>2$, the
density $\rho\left(  \lambda\right)  =0$ for $0<\lambda<\lambda_{c}$. In
contrast, in finite networks, Laplacian eigenvalues $\lambda_{i}$ exist in
this range, though only a very small fraction of the total number of the
eigenvalues. The statistics of this part of the spectrum determines the second
stage of the evolution of $p_{i0}\left(  t\right)  $ to the equilibrium. We
believe that this stage may be described in the framework of an approach
developed in Ref.~\cite{dms03} for calculation of intervertex distance
distributions. We leave this challenging problem for future study.

In summary, we have strictly shown that the region of low eigenvalues in the
Laplacian spectra of uncorrelated complex networks and the asymptotics of
random walks on them are essentially determined by the lowest vertex degree in
a network.

\begin{acknowledgments}
This work was partially supported by projects POCTI: FAT/46241/2002,
MAT/46176/2002, FIS/61665/2004, and BIA-BCM/62662/2004. S.N.D. and J.F.F.M.
were also supported by project DYSONET-NEST/012911. Authors thank A.~V.
Goltsev, B.~N. Shalaev, and M. Ostilli for useful discussions.
\end{acknowledgments}

\appendix 

\section{Other Laplacians and processes}

\label{a10000}

Three different forms of a Laplacian operator are discussed in literature. In
this paper we discussed the form (\ref{11}) corresponding to the process
defined by Eq.~(\ref{13}). The second form,%
\begin{equation}
L_{ij}=\delta_{ij}-\frac{1}{q_{i}}A_{ij}, \label{a10000-10}%
\end{equation}
corresponds to the following process:%
\begin{equation}
\dot{p}_{ij}\left(  t\right)  =\sum_{k=1}^{N}\frac{1}{q_{k}}A_{ik}%
p_{kj}(t)-p_{ij}(t),\ \ \ p_{ij}(0)=\delta_{ij}. \label{a10000-20}%
\end{equation}
This is a random walk process with the unit escape rate of a particle from any
vertex. The particle jumps to any of $q_{i}$ nearest neighbors of vertex $i$
with the same probability $1/q_{i}$. We do not consider this process here,
although it can be described in the framework of the approach of this article.
We have found that the singularity of the spectrum at the lowest eigenvalue of
this Laplacian and the long-time asymptotics of the autocorrelator of this
random walk are quite similar to those we found for the operator (\ref{11})
and the process (\ref{13}).

The third, \textquotedblleft normalized\textquotedblright, form,%
\begin{equation}
L_{ij}=\delta_{ij}-\frac{1}{\sqrt{q_{i}q_{j}}}A_{ij} \label{a10000-30}%
\end{equation}
(see, e.g., Ref.~\cite{clv03}) is, one may say, equivalent to the form
(\ref{a10000-20}) in the following sense. 
Operators (\ref{a10000-20}) and
(\ref{a10000-30}) are connected by a similarity transformation. The connecting operator $\hat{W}$ is diagonal: $W_{ij}=\delta_{ij}q_i$. 
These two operators have the same spectrum of eigenvalues. 
Their eigenfunctions are connected by the operator $\hat{W}$.

\section{Degree distribution in Z-representation}
\label{Z}

The Z-representation of a discrete random variable $q_{i}=0,1,2,\dots$ is
defined as
\begin{equation}
\varphi\left(  z\right)  =\frac{1}{N}\sum_{i=1}^{N}\left\langle z^{q_{i}%
}\right\rangle =\sum_{q=0}^{\infty}\Pi\left(  q\right)  z^{q}.\label{Z10}%
\end{equation}
$\varphi(z)$ is also called the generating function of $\Pi\left(  q\right)
$. It is obvious that $\varphi\left(  1\right)  =1$. Differentiating
$\varphi\left(  z\right)  $ and setting $z=1$, we obtain an expression for the
average vertex degree,
\begin{equation}
\varphi^{\prime}\left(  1\right)  =\sum_{q=0}^{\infty}q\Pi\left(  q\right)
=\left\langle q\right\rangle \equiv\bar{q}.\label{Z20}%
\end{equation}
In general,%
\begin{equation}
\left.  \left(  x\frac{d}{dz}\right)  ^{m}\varphi\left(  z\right)  \right\vert
_{x=1}=\sum_{q=0}^{\infty}q^{m}\Pi\left(  q\right)  =\left\langle
q^{m}\right\rangle .\label{Z30}%
\end{equation}

For branching numbers $b_{i}=q_{i}-1$ we have
\begin{equation}
\varphi_{1}\left(  z\right)  =\frac{1}{2L}\sum_{i,j=1}^{N}\left\langle
A_{ij}z^{b_{j}}\right\rangle =\frac{1}{2L}\sum_{j=1}^{N}\left\langle
q_{j}z^{q_{j}-1}\right\rangle =\frac{1}{\bar{q}}\sum_{q=0}^{\infty}q\Pi\left(
q\right)  z^{q-1}=\frac{\varphi^{\prime}\left(  z\right)  }{\bar{q}}.
\label{Z40}%
\end{equation}
The function $\varphi_{1}$ also obeys a normalization condition, $\varphi
_{1}\left(  1\right)  =1$.

This function was successfully used by Newman, Strogatz and Watts \cite{nsw01} 
(compare with the earlier works by Molloy and Reed, Ref.~\cite{mr95,mr98}) 
in their calculations of the size distributions of $n$-th connected components
of a vertex. Recall that this is a number of vertices which are not further
than $n$ steps from a vertex. For example, the distribution for the first
connected component in Z-representation is $z\varphi\left(  z\right)  $, for
the second one, it is $z\varphi\left[  z\varphi_{1}\left(  z\right)  \right]
$, and, in general, the distribution for an $n$-th component is $G_{n}\left(
z\right)  =z\varphi\left[  H_{n}\left(  z\right)  \right]  $. Here the
sequence $H_{n}$ is defined by the recursion relation 
\begin{equation}
H_{n}\left(  z\right)
=z\varphi_{1}\left[  H_{n-1}\left(  z\right)  \right], \ \ \ H_{0}\left(
z\right)  =z
. 
\label{Z45}
\end{equation}
Its stationary solution $H\left(  z\right)  $ satisfies the
equation:
\begin{equation}
H\left(  z\right)  =z\varphi_{1}\left[  H\left(  z\right)  \right]  .
\label{Z50}%
\end{equation}
So $G\left(  z\right)  =z\varphi\left[  H\left(  z\right)  \right]
=\left\langle z^{M_{i}}\right\rangle $ is the transformed probability function
that a randomly chosen vertex is in a connected component of size $M_{i}$.

The function $G(z)$ allows one to find, in particular, the relative size of a
giant connected component, $m_{\infty}=N_{\infty}/N$. Let us consider the
solutions of Eq. (\ref{Z50}) as $z\rightarrow+0$, $H\left(  +0\right)  =t_{c}%
$; $t_{c}=\varphi_{1}\left(  t_{c}\right)  $. Beside the trivial solution 
equal to zero, there is another solution, $t_{c}<1$ (see Fig.~\ref{phi1}):%
\begin{equation}
t_{c}=\varphi_{1}\left(  t_{c}\right)  ,\ \ 0\leq t_{c}<1,\ \ \text{if }%
z_{1}=\varphi_{1}^{\prime}\left(  1\right)  =\frac{\varphi^{\prime\prime
}\left(  1\right)  }{\varphi^{\prime}\left(  1\right)  }>1. \label{Z60}%
\end{equation}
So $\varphi\left(  t_{c}\right)  $ is the total relative size of all connected
components of the network, and the relative size of the giant connected component is
\begin{equation}
m_{\infty}=1-\varphi\left(  t_{c}\right)  . \label{Z70}%
\end{equation}
Note that the condition (\ref{Z60}) may be written as
\[
\sum_{q}q\left(  q-2\right)  \Pi\left(  q\right)  >0.
\]
If there are no \textquotedblleft dead ends\textquotedblright\ in the network,
then $t_{c}=\varphi\left(  t_{c}\right)  =0$, and almost all vertices in the
network are in the giant connected component.

\section{Equation for the distribution of $\tau$ and autocorrelator}

\label{T}

If $n>1$, Eq. (\ref{110}) may be written as (see Fig. \ref{tree})
\begin{equation}
sP_{n+1,i}\left(  s\right)  -\left[  P_{n,j}\left(  s\right)  -P_{n+1,i}%
\left(  s\right)  \right]  +\sum_{k=1}^{b_{n+1,i}}\left[  P_{n+1,i}\left(
s\right)  -P_{n+2,k}\left(  s\right)  \right]  =0 . \label{T10}%
\end{equation}
Dividing both parts of the equation by $P_{n+1,i}\left(  s\right)  $, and
taking into account the definition (\ref{120}), we get
\begin{equation}
\frac{\tau_{n,ij}\left(  s\right)  }{1-\tau_{n,ij}\left(  s\right)  }%
=s+\sum_{k=1}^{b_{n+1,i}}\tau_{n+1,jk}\left(  s\right)  . \label{T20}%
\end{equation}
If $n=0$, Eq. (\ref{110}) takes the form:
\begin{equation}
sP_{0}\left(  s\right)  +\sum_{k=1}^{q_{0}}\left[  P_{0}\left(  s\right)
-P_{1k}\left(  s\right)  \right]  =1 . \label{T30}%
\end{equation}
Dividing both sides of Eq. (\ref{T30}) by $P_{0}$, and taking into account Eq.
(\ref{120}), we obtain
\begin{equation}
P_{0}\left(  s\right)  =\left[  s+\sum_{k=1}^{q_{0}}\tau_{0,0k}\right]  ^{-1}
. \label{T40}%
\end{equation}

Recursive relations (\ref{T20}) express the set of random variables
$\tau_{n,ij}$, $n\geq0$, in terms of the set of independent and statistically
equivalent random variables $q_{m,i}$, $m>n$. It is important that the
variable $\tau_{n,ij}$ depends only on the degrees of vertices belonging to
the tree branch, which grows from the edge $\left(  n,i\right)  -\left(
n+1,j\right)  $. So in Eq. (\ref{20}), $\tau_{n,ij}\left(  s\right)  $ is
expressed through $q_{n+1,j}$ independent random variables: the branching
number $b_{n+1,j}=q_{n+1,j}-1$ and $b_{n+1,j}$ statistically equivalent
variables $\tau_{n+1,jk}$, $k=1,\dots b_{n+1,j}$. In the thermodynamic limit,
the statistical properties of branches, starting at any distance from the
initial vertex, are the same. Consequently, all random variables $\tau_{n,ij}$
are distributed equally, independently of $i,j$ and $n$. Then, omitting
unnecessary indices, one can rewrite Eq. (\ref{T20}) as%
\begin{equation}
e^{y}\exp\left[  -\frac{y}{1-\tau\left(  s\right)  }\right]  =e^{-sy}%
\prod_{k=1}^{b}e^{-y\tau_{k}\left(  s\right)  }.\label{T50}%
\end{equation}
The next step is averaging both the parts of Eq. (\ref{T40}). We use 
definition (\ref{130}), properties of statistical equivalence of $\tau$, and
mutual independence of the branching number $b$ and all $\tau_{k}$. We also
use the following integral identity:%
\begin{equation}
e^{-y/\alpha}=\frac{\sqrt{y}}{i\pi}\int_{-i\infty+\delta}^{+i\infty+\delta
}\frac{dx}{\sqrt{x}}K_{1}\left(  2\sqrt{xy}\right)  e^{\alpha x},\label{T60}%
\end{equation}
where $K_{1}$ is the MacDonald function of index $1$. Then
\[
\exp\left(  -\frac{y}{1-\tau}\right)  =\frac{\sqrt{y}}{i\pi}%
\int_{-i\infty+\delta}^{+i\infty+\delta}\frac{dx}{\sqrt{x}}K_{1}\left(
2\sqrt{xy}\right)  e^{\left(  1-\tau\right)  x}.
\]
Finally, we have
\begin{equation}
e^{y}\frac{\sqrt{y}}{i\pi}\int_{-i\infty+\delta}^{+i\infty+\delta}\frac
{dx}{\sqrt{x}}K_{1}\left(  2\sqrt{xy}\right)  e^{x}T(s,x)
=e^{-sy}\left\langle
\left\{  \left[  T\left(  s,y\right)  \right]  ^{b}\right\}  \right\rangle
=e^{-sy}\varphi_{1}\left[  T\left(  s,y\right)  \right]  ,\label{T70}%
\end{equation}
where definition (\ref{Z40}) was used. (Here $b$ is a branching
coefficient of some edge.)

Introducing $\xi\left(  s\right)  =\tau\left(  s\right)  /\left[
1-\tau\left(  s\right)  \right]  $, so that $\tau\left(  s\right)  =\xi\left(
s\right)  /\left[  1+\xi\left(  s\right)  \right]  $, enables us to use the
following integral identity equivalent to Eq. (\ref{T60}):%
\begin{equation}
e^{x/\alpha}=1+\sqrt{x}\int_{0}^{\infty}\frac{dy}{\sqrt{y}}I_{1}\left(
2\sqrt{xy}\right)  e^{-\alpha y}. \label{T80}%
\end{equation}
Here $I_{1}$ is the modified Bessel function of index $1$. So
\[
e^{x}\left\langle e^{-x\tau\left(  s\right)  }\right\rangle =1+\sqrt{x}%
\int_{0}^{\infty}\frac{dy}{\sqrt{y}}I_{1}\left(  2\sqrt{xy}\right)
e^{-y}\left\langle e^{-s\xi\left(  x\right)  }\right\rangle .
\]
Note Eq. (\ref{T20}) leads to the relation:
\begin{equation}
\xi\left(  s\right)  =s+\sum_{k=1}^{b}\tau_{k}\left(  s\right)  . \label{T85}%
\end{equation}
Averaging in the integral in the same way as in Eq. (\ref{T70}), we arrive at
Eq. (\ref{140}).

After the averaging, Eq. (\ref{T40}) takes the form:%
\[
\bar{P}_{0}\left(  s\right)  =\int_{0}^{\infty}dx~e^{-sx}\left\langle
\prod_{k=1}^{q_{0}}\exp\left[  -x\tau_{0,0k}\right]  \right\rangle .
\]
Taking into account the property of statistical independence and equivalence 
indicated above, we obtain Eq. (\ref{150}).

\section{Derivation of the recursion relation}
\label{U}

Let us consider the following expression:
\begin{equation}
\frac{S_{n,ij}^{\left(  l\right)  }\left(  s\right)  }{1-\tau_{n,ij}\left(
s\right)  }\exp\left[  -\frac{y\tau_{n,ij}\left(  s\right)  }{1-\tau
_{n,ij}\left(  s\right)  }\right]  . \label{U10}%
\end{equation}
Differentiating the identity (\ref{T60}) with respect to $x$ we have
\begin{equation}
\frac{1}{\alpha}e^{-y/\alpha}=\int_{-i\infty+\delta}^{+i\infty+\delta}\frac{dx}{\pi i}%
K_{0}\left(  2\sqrt{xy}\right)  \alpha e^{\alpha x}. \label{U20}%
\end{equation}
Substituting $\alpha=1/\left(  1-\tau\right)  $ and using the definition of
$U_{l}$, Eq. (\ref{170}), we transform the expression (\ref{U10}) into
\begin{equation}
e^{y}\int_{-i\infty+\delta}^{+i\infty+\delta}\frac{dx}{\pi i}K_{0}\left(
2\sqrt{xy}\right)  S_{n,ij} \exp\left\{  x\left[  1-\tau_{n,ij}(s)
\right]  \right\}  , \label{U30}%
\end{equation}
where $K_{0}$ is MacDonald's function of index $0$. In the infinite network
the expression (\ref{U30}) is independent of the chosen edge $\left(
n,i\right)  -\left(  n+1,j\right)  $ and depends only on $l$ and on $s$.
Averaging Eq. (\ref{U30}) we get
\begin{equation}
e^{y}\int_{-i\infty+\delta}^{+i\infty+\delta}\frac{dx}{\pi i}K_{0}\left(
2\sqrt{xy}\right)  e^{x}U_{l}\left(  s,x\right)  . \label{U35}%
\end{equation}

On the other hand, due to the tree-like structure, the first multiplier in the
angular brackets in Eq. (\ref{U10}) may be expressed as a sum of terms with
$l\rightarrow l-1$ (see Fig.~\ref{tree}):%
\begin{equation}
S_{n,;i,j}^{\left(  l\right)  }\left(  s\right)  =\left[  1-\tau_{n,ij}\left(
s\right)  \right]  \sum_{k=1}^{b_{n+1,j}}S_{n+1,jk}^{\left(  l-1\right)
}\left(  s\right)  . \label{U40}%
\end{equation}
Using Eq. (\ref{40}) together with Eq. (\ref{T20}), we see that the expression
(\ref{U10}) is equal to the following one:
\begin{equation}
e^{-sy}\prod_{k=1}^{b_{n+1,i}}\exp[-y\tau_{n+1,jk}\left(  s\right)
]\sum_{m=1}^{b_{n+1,j}}S_{n+1,jm}^{\left(  l-1\right)  }\left(  s\right)  .
\label{U50}%
\end{equation}
Let us average Eq. (\ref{U50}) taking into account the statistical properties
of the variables $\tau$ and $b$ (or $q$), indicated above. Note that in each
of $b_{n+1,j}$ terms we have $b_{n+1,j}-1$ multipliers $\left\langle \exp
[-y\tau_{n+1,jk}\left(  s\right)  ]\right\rangle =T\left(  s,y\right)  $ with
$k\neq m$, and the multiplier $\left\langle S_{n+1,jm}^{\left(  l-1\right)
}\left(  s\right)  \exp[-y\tau_{n+1,jk}\left(  s\right)  ]\right\rangle
=U_{l-1}\left(  s,y\right)  $. So the remaining average 
over $b_{n+1,j}\equiv b$
can be easily performed, which gives%
\begin{equation}
e^{-sy}
\sum_b b\left[  T\left(  s,y\right)  \right]  ^{b-1} \Pi_1(b) 
U_{l-1}(s,y)=e^{-sy}\varphi_{1}^{\prime}\left[  T\left(  s,y\right)  \right]
U_{l-1}\left(  s,y\right)  . \label{U60}%
\end{equation}
[Recall that the distribution function of $b$ is $\Pi_{1}\left(  b\right)  $,
Eq. (\ref{115}), i.e., $\varphi_{1}\left(  z\right)  $ in Z-representation,
Eq. (\ref{Z40}).] Equating expression (\ref{U35}) to Eq. (\ref{60}), that is a
different representation of expression (\ref{U10}), we derive the recursion
relation for $U_{n}$ in the following form:
\begin{equation}
e^{y}\int_{-i\infty+\delta}^{+i\infty+\delta}\frac{dx}{\pi i}K_{0}\left(
2\sqrt{xy}\right)  e^{x}U_{l}\left(  s,x\right)  =e^{-sy}\varphi_{1}^{\prime
}\left[  T\left(  s,y\right)  \right]  U_{l-1}\left(  s,y\right)  .
\label{U70}%
\end{equation}
Equation (\ref{2020}) for eigenfunctions $\psi_{m}\left(  s\right)  $ can also
be written as
\begin{equation}
e^{y}\int_{-i\infty+\delta}^{+i\infty+\delta}\frac{dx}{\pi i}K_{0}\left(
2\sqrt{xy}\right)  e^{x}\psi_{m}\left(  s,x\right)  =e^{-sy}\varphi
_{1}^{\prime}\left[  T\left(  s,y\right)  \right]  \psi_{m}\left(  s,y\right)
. \label{U75}%
\end{equation}

Let us now replace $\tau$ in the definition of $U_{l}$, Eq.~(\ref{170}), with
its expression in terms of the random variable $\xi\left(  s\right)  $,
$\tau\left(  s\right)  =\xi\left(  s\right)  /\left[  1+\xi\left(  s\right)
\right]  $. In turn, for $\xi\left(  s\right)  $ we use relation (\ref{T85}).
Again, use Eq. (\ref{U40}) for $S$. Differentiating integral identity
(\ref{T80}) with respect to $x$ gives
\begin{equation}
e^{x/\alpha}=\int_0^\infty dy I_{0}\left(2\sqrt{xy}\right) \alpha  e^{-\alpha y}. \label{U80}%
\end{equation}
Then we have
\begin{eqnarray}
&&
U_{l}(s,x) 
= e^{-x}\int_0^\infty dy I_{0}\left(2\sqrt{xy}\right) e^{-(1+s)y} 
\left\langle 
\prod_{k=1}^{b_{n+1,i}} 
e^{-y\tau_{n+1,jk}(s)}\sum_{m=1}^{b_{n+1,j}} S_{n+1,jm}^{(l-1)}
\right\rangle  
\nonumber
\\[5pt]
&&
\ \ \ \ \ \ \ \ \ \ \ 
= e^{-x} \int_0^\infty dy I_{0}\left(2\sqrt{xy}\right) e^{-(1+s)y} 
\sum_b b [T(s,y)]^b \Pi_1(b) U_{l-1}(s,y) 
. 
\label{U90}
\end{eqnarray}
Using Eq. (\ref{U60}) for averaging over $b$ readily leads to Eq.~(\ref{180}). 
The initial condition (\ref{190}) follows directly from the definition of
$U_{l}$, Eq. (\ref{170}).

\section{Asymptotic solutions of integral equations}

\label{A}

Calculating the asymptotics at large $x$ we replace the MacDonald functions
$K_{\nu}\left(  2\sqrt{xy}\right)  $ with the leading term of its asymptotic
expression:%
\begin{equation}
K_{\nu}\left(  z\right)  \rightarrow\sqrt{\frac{\pi}{2z}}e^{-z}. \label{A10}%
\end{equation}
This asymptotics is independent of $\nu$. Then Eqs. (\ref{T70}) and
(\ref{U70}) at large $x$ take the forms:
\begin{equation}
e^{y}\frac{y^{1/4}}{i\sqrt{2\pi}}\int_{-i\infty+\delta}^{+i\infty+\delta}%
\frac{dx}{x^{3/4}}\exp\left(  x-2\sqrt{xy}\right)  T\left(  s,x\right)
=e^{-sy}\varphi_{1}\left[  T\left(  s,y\right)  \right]  \label{A20}%
\end{equation}
and
\begin{equation}
e^{y}\frac{1}{iy^{1/4}\sqrt{2\pi}}\int_{-i\infty+\delta}^{+i\infty+\delta
}\frac{dx}{x^{1/4}}\exp\left(  x-2\sqrt{xy}\right)  U_{l}\left(  s,x\right)
=e^{-sy}\varphi_{1}^{\prime}\left[  T\left(  s,y\right)  \right]
U_{l-1}\left(  s,y\right)  . \label{A30}%
\end{equation}
Equation (\ref{U75}) in the asymptotic limit has the form:%
\begin{equation}
e^{y}\frac{1}{iy^{1/4}\sqrt{2\pi}}\int_{-i\infty+\delta}^{+i\infty+\delta
}\frac{dx}{x^{1/4}}\exp\left(  x-2\sqrt{xy}\right)  \psi_{m}\left(
s,x\right)  =\mu_{m}\left(  s\right)  e^{-sy}\varphi_{1}^{\prime}\left[
T\left(  s,y\right)  \right]  \psi_{m}\left(  s,y\right)  . \label{A40}%
\end{equation}

Let us first consider Eq. (\ref{A20}) in the case $s>0$, $y\rightarrow\infty$.
According to the definition (\ref{130}) of $T\left(  s,x\right)  $, this
function is the Laplace transform of the probability density of a random
variable $\tau\left(  s\right)  $. This variable satisfies the condition
$0<\tau<1$, as it follows e.g., from the recursion relation (\ref{T20}).
Hence, (i) the function $T\left(  s,x\right)  $ is analytic everywhere in the
complex plane $x$, and $T\left(  s,x\right)  \rightarrow0$ as
$\operatorname{Re}x\rightarrow\infty$; (ii) $T\left(  s,x\right)  $ cannot
decrease with $x$ faster than exponentially. Then $T\left(  s,x\right)  $ can
represented as
\begin{equation}
T\left(  s,x\right)  =A\exp\left[  -\tau_{m}\left(  s\right)  x-\vartheta
_{0}\left(  s,x\right)  \right]  , \label{A50}%
\end{equation}
where $\tau_{m}\geq0$, and $\partial_{x}\vartheta_{0}\left(  s,x\right)
\rightarrow0$ as $x\rightarrow+\infty$. If $q_{m}>1$ (we will consider this
case separately), then $\varphi_{1}$ on the right-hand side of Eq.~(\ref{A20})
can be replaced with its leading term, $\varphi_{1}\left(  z\right)
\rightarrow q_{m}\Pi\left(  q_{m}\right)  z^{q_{m}-1}$:%
\begin{equation}
\frac{y^{1/4}}{2i\pi^{1/2}}\!\int_{-i\infty}^{+i\infty}\!\frac{dx}{x^{3/4}%
}\exp\left[  (1{-}\tau_{m})x-2\sqrt{xy}-\vartheta_{0}(x)\right]  =\frac{q_{m}%
}{\bar{q}}\Pi(q_{m})A^{q_{m}-2}\exp\left[  -(1{+}s)y-(q_{m}{-}1)\tau
_{m}y-(q_{m}{-}1)\vartheta_{0}(y)\right]  . \label{A60}%
\end{equation}
The integral on the left-hand side may be treated in the saddle point
approximation. The saddle point equation is the condition that the derivative
of the function in the exponent becomes equal zero, namely,
\begin{equation}
y=x\left[  1-\tau_{m}-\vartheta_{0}^{\prime}\left(  x\right)  \right]  ^{2}.
\label{A70}%
\end{equation}
This equation also expresses $y$ in terms of $x$. On the right-hand side, we
assume that
\[
\vartheta_{0}\left(  y\right)  \approx\vartheta_{0}\left[  \left(  1-\tau
_{m}\right)  ^{2}x\right]  -2\left(  1-\tau_{m}\right)  x\vartheta_{0}%
^{\prime}\left(  x\right)  \vartheta_{0}^{\prime}\left[  \left(  1-\tau
_{m}\right)  ^{2}x\right]  .
\]
One must prove afterwards that the neglected terms of the order of
$x\vartheta_{0}^{\prime3}$ and with higher derivatives are small. If we set
$\vartheta_{0}^{\prime}=0$ in the pre-exponential factor of the saddle point
approximation, it reduces to $1$. So we arrive at the following equation for
$\vartheta_{0}$:%
\begin{align}
&  \!\!\!\!\!\!\ln\left[  \frac{q_{m}}{\bar{q}}\Pi
(q_{m})A^{q_{m}-2}\right]  +(1{-}\tau_{m})^{2}\left[  s+(q_{m}{-}1)\tau
_{m}-\frac{\tau_{m}}{1-\tau_{m}}\right]  x-2(1{-}\tau_{m})\left[  s+(q_{m}%
{-}1)\tau_{m}-\frac{\tau_{m}}{1{-}\tau_{m}}\right]  x\vartheta_{0}^{\prime
}(x)\nonumber\\[5pt]
&  \!\!\!\!\!\!+(q_{m}{-}1)\vartheta_{0}[(1-\tau_{m}%
)^{2}x]-\vartheta_{0}(x)+[1+s+(q_{m}{-}1)\tau_{m}]x\vartheta_{0}^{\prime
2}(x)-2(q_{m}{-}1)(1-\tau_{m})x\vartheta_{0}^{\prime}(x)\vartheta_{0}^{\prime
}[(1-\tau_{m})^{2}x]=0. 
\label{A80}%
\end{align}
The main term of this equation, linear in $x$, reduces to zero if
\begin{equation}
\frac{\tau_{m}}{1-\tau_{m}}=s+\left(  q_{m}-1\right)  \tau_{m}. \label{A90}%
\end{equation}
It also reduces the third term in Eq. (\ref{A80}) to zero. Suppose that our
network is a the regular Bethe lattice with the coordination number $q_{m}$.
Then Eq. (\ref{T70}), equivalent to Eq. (\ref{150}), has the exact solution
$T\left(  s,x\right)  =\exp\left[  -\tau_{m}\left(  s\right)  x\right]  $,
where $\tau_{m}\left(  s\right)  $ is the proper solution of Eq. (\ref{A90}).
This $\tau_{m}\left(  s\right)  $ is a regular function of $s$ as
$s>s_{c}=-q_{m}+2\sqrt{q_{m}-1}$, $s_{c}<0$, and $\tau_{m}\left(
s_{c}\right)  \equiv\tau_{c}=1-1/\sqrt{q_{m}-1}$. At $s=s_{c}$, $\tau
_{m}\left(  s\right)  $ has a square root singularity. So in the regular Bethe
lattice, the density of Laplacian eigenvalues $\rho\left(  \lambda\right)  $ is
nonzero at $\lambda>\lambda_{c}=-s_{c}=q_{m}-2\sqrt{q_{m}-1}$, and
$\rho\left(  \lambda\right)  \sim\sqrt{\lambda-\lambda_{c}}$ at $\lambda
-\lambda_{c}\ll1$. Thus, we can conclude, that for any network with $q_{m}>1$,
the edge of the spectrum is $\lambda_{c}\geq0$. Moreover, $\lambda_{c}%
(q_{m}>2)>0$. In random networks, the asymptotics of $\rho(\lambda)$ turn out
to be sharply different from a regular Bethe lattice.

Requiring that the main correction to the leading term in Eq. (\ref{A90}) also
asymptotically vanish gives $\left(  q_{m}-1\right)  \vartheta_{0}\left[
\left(  1-\tau_{m}\right)  ^{2}x\right]  =\vartheta_{0}\left(  x\right)  $.
This equality is satisfied when
\begin{equation}
\vartheta_{0}\left(  x\right)  =Bx^{\alpha},\ \ \alpha=\frac{\ln\left(
q_{m}-1\right)  }{2\ln\left[  1/\left(  1-\tau_{m}\right)  \right]  }%
=\frac{\ln\left[  1/\left(  1-\tau_{c}\right)  \right]  }{\ln\left[  1/\left(
1-\tau_{m}\right)  \right]  }. \label{A100}%
\end{equation}
If $s>s_{c}$, then $\tau_{m}>\tau_{c}$ and $\alpha<1$. This means that all 
approximations made during the derivation are justified. Therefore the
integral in Eq. (\ref{150}) is convergent, and $\bar{P}_{0}\left(  s\right)  $
is a regular function of $s$. If, however, $s\rightarrow s_{c}$, then
$\alpha\rightarrow1$. That is, the last two terms in Eq. (\ref{A80}) should be
also taken into account when $s$ is close to $s_{c}$. In this region of $s$,
Eq.~(\ref{A80}) and a similar equation for the asymptotics of $\psi
_{0}\left(  x\right)  $, which can be derived from Eq. (\ref{U75}), must be
treated in different ways for $q_{m}>2$ and for $q_{m}=2$. Note that if
$q_{m}=1$, then Eq. (\ref{A80}) must be replaced with a slightly different equation.

\subsection{Minimum degree $q_{m}>2$}

\label{A3}

In this case we can set the first term in Eq. (\ref{A80}) to $0$, properly
choosing the value of the constant $A$ in Eq. (\ref{A50}). We set $T\left(
s,x\right)  =\exp\left[  -\vartheta\left(  \delta,x\right)  \right]  $. Here
$\vartheta\left(  \delta,x\right)  =\vartheta_{0}+\tau_{m}x$ includes,
besides $\vartheta_{0}$, also a slowly varying linear term $\tau_{m}x$. Here we
introduce a small variable $\delta=\sqrt{s-s_{c}}$.

First, let us consider Eq. (\ref{A80}) at $s=s_{c}$ and $\tau=\tau_{c}$. We
have
\[
\left(  q_{m}-1\right)  \vartheta\left(  \frac{x}{q_{m}-1}\right)
-\vartheta\left(  x\right)  +\sqrt{q_{m}-1}y\vartheta^{\prime2}\left(
x\right)  -2\sqrt{q_{m}-1}y\vartheta^{\prime}\left(  x\right)  \vartheta
^{\prime}\left(  \frac{x}{q_{m}-1}\right)  =0.
\]
Now we make the substitution: $\vartheta\left(  x\right)  =\left(
q_{m}-1\right)  ^{-1/2}x\chi\left(  \ln x\right)  $. We assume that $\chi$ is
a small and slowly varying function of its argument. Then we make the
following approximations, which must be justified afterwards. Replace
$\chi\left[  z-\ln\left(  q_{m}-1\right)  \right]  $ in the first term with $\chi\left(  z\right)  -\chi^{\prime}\left(  z\right)  \ln\left(
q_{m}-1\right)  $, where $z=\ln
y$, and neglect all derivatives of $\chi$ in the last two
terms. As a result we get
\[
\ln\left(  q_{m}-1\right)  \chi^{\prime}\left(  z\right)  +\chi^{2}\left(
z\right)  =0.
\]
This equation has the solution: $\chi\left(  z\right)  =\ln\left(
q_{m}-1\right)  /\left(  z+c\right)  $, where $c\sim1$ is some constant of
integration. Thus, finally, we obtain
\begin{equation}
\vartheta\left(  \delta=0,x\right)  =\frac{x\ln\left(  q_{m}-1\right)  }%
{\sqrt{q_{m}-1}\ln\left(  Cx\right)  }. \label{A110}%
\end{equation}

Now assume $\left\vert \delta\right\vert ^{2}=\left\vert s-s_{c}=\left\vert
\lambda-\lambda_{c}\right\vert \right\vert \ll1$. The first term in Eq.
(\ref{A80}) reduces to $\delta^{2}y/\left(  q_{m}-1\right)  $. We neglect the
second term of the equation, assuming it to be small. After the same set of
substitutions and approximations as in the case $s=s_{c}$, we have the
following equation for $\chi\left(  z\right)  $:%
\[
\ln\left(  q_{m}-1\right)  \chi^{\prime}\left(  z\right)  +\chi^{2}\left(
z\right)  =\frac{\delta^{2}}{\sqrt{q_{m}-1}}.
\]
Solving this equation, we obtain the following result for $\vartheta\left(
\delta,x\right)  $:%
\begin{equation}
\vartheta\left(  \delta,x\right)  =\frac{\delta x}{\left(  q_{m}-1\right)
^{3/4}}\coth\left[  \frac{\delta\ln\left(  Cx\right)  }{\left(  q_{m}%
-1\right)  ^{1/4}\ln\left(  q_{m}-1\right)  }\right]  . \label{A120}%
\end{equation}
After substitution $\delta=\sqrt{s-s_{c}}$, this turns into Eq. (\ref{350}).

\subsection{Minimum degree $q_{m}=2$}

\label{A2}

Here $\tau_{m}\left(  s\right)  \rightarrow0$ as $s\rightarrow0$. Then at
small $\left\vert s\right\vert $ we can consider $\vartheta\left(  s,x\right)
=\vartheta_{0}\left(  s,x\right)  +\tau_{m}\left(  s\right)  x$ as a slowly
varying function. When calculating the integral in Eq. (\ref{A20}) in the
saddle point approximation, we take also into account the pre-exponential
factor as a correction, though it is close to $1$. Replacing on the
right-hand side $\varphi_{1}\left(  T\right)  $ with its leading term, linear
on $T$, and taking into account the saddle point equation $y=x\left[
1-\vartheta^{\prime}\left(  x\right)  \right]  ^{2}$, we have
\begin{equation}
\left[  \frac{1-\vartheta^{\prime}\left(  x\right)  }{1-\vartheta^{\prime
}\left(  x\right)  -2x\vartheta^{\prime\prime}\left(  x\right)  }\right]
^{1/2}\exp\left[  x\vartheta^{\prime2}\left(  x\right)  -\vartheta\left(
x\right)  \right]  =\frac{2\Pi\left(  2\right)  }{\bar{q}}\exp\left[
-sx-\vartheta\left(  x\right)  +2x\vartheta^{\prime2}\left(  x\right)
\right]  . \label{A210}%
\end{equation}
Here we omitted negligibly small terms: $2sx\vartheta^{\prime}\left(
x\right)  $ and others.

Accounting for the smallness of $\vartheta^{\prime}$ and $x\vartheta
^{\prime\prime}$, we obtain the equation:
\begin{equation}
x\vartheta^{\prime2}\left(  x\right)  -y\vartheta^{\prime\prime}\left(
x\right)  =sx+\frac{a}{\pi},\ \ a=\pi\ln\left[  \frac{\bar{q}}{2\Pi\left(
2\right)  }\right]  >0. \label{A220}%
\end{equation}
We assume that the first term on the left-hand side is small and search for
the solution of this equation in the form $\vartheta=\vartheta_{1}%
+\vartheta_{2}$. Here $\vartheta_{1}$ must be found from $x\vartheta
_{1}^{\prime2}=sx+a/\pi$. At $s=0$ we find $\vartheta_{1}=2\sqrt{ax/\pi}+c$,
where $C$ is some constant of integration, $C\sim1$. At $s\neq0$, performing
the integration, we have
\begin{equation}
\vartheta_{1}\left(  x\right)  =\frac{1}{\sqrt{s}}f\left(  sx\right)
+C,\ \ f\left(  z\right)  =\sqrt{z\left(  a/\pi+z\right)  }+\frac{a}{\pi
}\operatorname{arcsinh}\sqrt{\frac{\pi z}{a}}. \label{A230}%
\end{equation}
In principle, here $C=C\left(  s\right)  $, but for small $s$ one can set
$C\left(  s\right)  =C=C\left(  0\right)  $. For $\vartheta_{2}$ we have
$2\vartheta_{1}^{\prime}\vartheta_{2}^{\prime}=\vartheta_{1}^{\prime\prime}$.
Therefore up to the constant, $\vartheta_{2}=\left(  \ln\vartheta_{1}^{\prime
}\right)  /2$. As a result, we have asymptotically the expression (\ref{460})
for $\vartheta=\vartheta_{1}+\vartheta_{2}$.

We replace in Eq. (\ref{U75}) $K_{0}$ with its asymptotic (\ref{A10}) at large
values of argument. Then, taking into account that $T\left(  s,x\right)  $ is
small at large $x$, we replace $\varphi_{1}\left(  T\right)  $ on the right-hand side with its value of zero argument, $2\Pi\left(  2\right)  /\bar{q}$.
As a result, we arrive at the following equation for $\psi_{0}\left(
s,x\right)  =\exp\left[  -\varkappa\left(  s,x\right)  \right]  $:%
\begin{equation}
\frac{e^{y}}{2iy^{1/4}\sqrt{\pi}}\int_{-i\infty+\delta}^{+i\infty+\delta}%
\frac{dx}{x^{1/4}}\exp\left[  -2\sqrt{xy}+x-\varkappa\left(  x\right)
\right]  =\mu_{0}\left(  s\right)  \frac{2}{\bar{q}}\Pi\left(  2\right)
\exp\left[  -sy-\varkappa\left(  y\right)  \right]  . \label{A240}%
\end{equation}
This equation differs from Eq. (\ref{A60}) for $\vartheta\left(  x\right)
=\vartheta_{0}\left(  x\right)  +\tau_{m}\left(  s\right)  x$ only in the
pre-exponential factor on the left-hand side. Quite analogously to Eq.
(\ref{A220}), we obtain an equation for $\varkappa$,
\begin{equation}
x\varkappa^{\prime2}\left(  x\right)  -x\varkappa^{\prime\prime}\left(
x\right)  -\varkappa^{\prime}\left(  x\right)  =sy+\frac{a^{\prime}}{\pi
}, \ \ a^{\prime}=\pi\ln\left[  \frac{\mu_{0}\left(  s\right)  \bar{q}}%
{2\Pi\left(  2\right)  }\right]  \approx a>0. \label{A250}%
\end{equation}
Here $b$ is given in Eq. (\ref{A220}). Comparing the above equation with Eq.
(\ref{A220}), we conclude that $\varkappa\left(  x\right)  =\vartheta
_{1}\left(  x\right)  +\varkappa_{2}\left(  x\right)  $, where $\vartheta
_{1}\left(  x\right)  $ is given by Eq. (\ref{A230}), and $\varkappa_{2}$ must
be found from $2x\vartheta_{1}^{\prime}\left(  x\right)  \varkappa_{2}%
^{\prime}\left(  x\right)  =x\vartheta_{1}^{\prime\prime}\left(  y\right)
+\vartheta_{1}^{\prime}\left(  x\right)  $. The solution is $\varkappa
_{2}=\left(  \ln x\vartheta_{1}^{\prime}\right)  /2$. As a result, accounting
for Eq. (\ref{A230}), we obtain the expression (\ref{482}) for $\psi_{0}%
=\exp\left(  -\varkappa\right)  $.

\subsection{Minimum degree $q_{m}=1$}

\label{A1}

To obtain equation for $\Theta\left(  z\right)  =\lim_{s\rightarrow0}T\left(
s,z/s\right)  $, let us start with Eq. (\ref{A20}), which is valid as
$y\rightarrow\infty$. Let us replace $y$ in this equation with $z/s$
simultaneously changing the integration variable: $x=\zeta/s$. Then we have
\begin{equation}
\frac{z^{1/4}}{i\sqrt{2\pi s}}\int_{-i\infty+\delta}^{+i\infty+\delta}%
\frac{d\zeta}{\zeta^{3/4}}\exp\left[  \frac{\left(  \sqrt{z}-\sqrt{\zeta
}\right)  ^{2}}{s}\right]  T\left(  s,\frac{\zeta}{s}\right)  =e^{-z}%
\varphi_{1}\left[  T\left(  s,z/s\right)  \right]  . \label{A310}%
\end{equation}
In the limit $s\rightarrow0$ the saddle point approximation becomes exact,
with the saddle point condition simply $\zeta_{c}=z$. So, assuming that the
limit (\ref{210}) of the function $T$ exists, we immediately arrive at Eq.~(\ref{220}) for $\Theta\left(  z\right)  $. The recursion relation
(\ref{260}) is obtained in the same way by using the asymptotic equation
(\ref{A30}).

We can reasonably assume that at small $s$ and large $x$,
\begin{equation}
T\left(  s,x\right)  =\Theta\left(  sx\right)  +\exp\left[  -\vartheta\left(
s,x\right)  \right]  , \label{A320}%
\end{equation}
where the last term is small. Substituting this into Eq. (\ref{A20}) and
linearizing the right-hand side with respect to $e^{-\vartheta}$, we obtain
\begin{equation}
\frac{e^{y}y^{1/4}}{2i\sqrt{\pi}}\int_{-i\infty+\delta}^{+i\infty+\delta}%
\frac{dx}{x^{3/4}}\exp\left[  -2\sqrt{xy}+x-\vartheta\left(  x\right)
\right]  =\varphi_{1}^{\prime}\left[  \Theta\left(  sy\right)  \right]
\exp\left[  -sy-\vartheta\left(  y\right)  \right]  . \label{A330}%
\end{equation}
Repeating the steps leading to Eq. (\ref{A220}), we get
\begin{equation}
x\vartheta^{\prime2}\left(  x\right)  -x\vartheta^{\prime\prime}\left(
x\right)  =sx-\ln\varphi_{1}^{\prime}\left[  \Theta\left(  sx\right)  \right]
. \label{A340}%
\end{equation}
Treating the second term on the left-hand side as a perturbation, we set
$\vartheta=\vartheta_{1}+\vartheta_{2}$, $\left\vert \vartheta_{2}\right\vert
\ll\left\vert \vartheta_{1}\right\vert $. The equations for $\vartheta_{1}$
and $\vartheta_{2}$ are
\[
x\vartheta_{1}^{\prime2}\left(  x\right)  =sx-\ln\varphi_{1}^{\prime}\left[
\Theta\left(  sx\right)  \right]  ,\ \ 2\vartheta_{1}^{\prime}\left(
x\right)  \vartheta_{2}^{\prime}\left(  x\right)  =\vartheta_{1}^{\prime
\prime}\left(  x\right)  .
\]
Then one can easily obtain
\begin{equation}
\vartheta_{1}\left(  s,x\right)  =\frac{1}{\sqrt{s}}g\left(  sx\right)
+C, \ \ g\left(  z\right)  =\int_{0}^{z}d\zeta\sqrt{1-\frac{\ln\varphi_{1}%
^{\prime}\left[  \Theta\left(  \zeta\right)  \right]  }{\zeta}}, \ \ \vartheta
_{2}\left(  s,x\right)  =\frac{1}{4}\ln\left[  sg^{\prime2}\left(  sx\right)
\right]  , \label{A350}%
\end{equation}
where $C\sim1$. Finally, for $\vartheta\left(  x\right)  =\vartheta_{1}\left(
x\right)  +\vartheta_{2}\left(  x\right)  $ we get formula (\ref{500}).

The function $g\left(  z\right)  $ has a singularity point at $z=z_{s}<0$,
when the expression in the square root under the integral becomes $0$, i.e.
when $\varphi_{1}^{\prime}\left[  \Theta\left(  z_{s}\right)  \right]
=e^{z_{s}}$. Here $z_{s}$ is defined as $t_{s}=\Theta\left(  e^{-z_{s}%
}\right)  $, $t_{s}$ is the solution of Eq. (\ref{510}) [see Fig.
(\ref{phi1})]. It is obvious that $g\left(  z_{s}\right)  \equiv ia$ is
imaginary, because the expression in the square root in Eq. (\ref{A350}) is
negative. The calculation of $a$ may be simplified if we replace the
integration variable $\zeta$ with $\xi=\Theta\left(  \zeta\right)  $. We use
the definition of the function $\Theta$, Eq. (\ref{220}), from which 
$\zeta=\ln\left[  \varphi_{1}\left(  \xi\right)  /\xi\right]  $ follows, so
$d\zeta=\left[  \varphi_{1}^{\prime}\left(  \xi\right)  /\varphi_{1}\left(
\xi\right)  -1/\xi\right]  d\xi$. We substitute these relations into the
integral for $g\left(  z_{s}\right)  $ in Eq. (\ref{A350}), and take into
account that $\xi=t_{c}=\Theta\left(  0\right)  $ at the lower limit of
integration, $\zeta=0$ while $\xi=t_{s}$ on the upper limit. This gives Eq.
(\ref{530}). Note that $z=t_{s}$ is a singularity point of the function
$\Theta\left(  z\right)  $. Since the derivative of the reverse function
$z=\ln\left[  \varphi_{1}\left(  \Theta\right)  /\Theta\right]  $ is zero,
$\Theta\left(  z_{s}+\eta\right)  =t_{s}+\mathcal{O}\left(  \sqrt{\eta
}\right)  $ at small $\left\vert \eta\right\vert $. Therefore in Eq.
(\ref{A350}) $\ln\varphi_{1}^{\prime}\left[  \Theta\left(  z_{s}+\eta\right)
\right]  /\left(  z_{s}+\eta\right)  -1\sim\sqrt{\eta}$ and so $g\left(
z_{s}+\eta\right)  =ia+\mathcal{O}\left(  \eta^{5/4}\right)  $.

The calculation of the asymptotics of the eigenfunction $\psi_{0}\left(
s,x\right)  $ is quite similar to that for $q_{m}=2$. Let us represent
$\psi_{0}\left(  s,x\right)  =\exp\left[  -\varkappa\left(  s,x\right)
\right]  $. From Eq.~(\ref{U75}) we obtain the integral equation for
$\varkappa$ which differs from Eq.~(\ref{A240}) only in that the constant
$\varphi_{1}^{\prime}\left(  0\right)  =2\Pi\left(  2\right)  /\bar{q}$ should
be replaced with the function $\varphi_{1}^{\prime}\left[  \Theta\left(
sy\right)  \right]  $. Proceeding further, we have
\[
x\varkappa^{\prime2}\left(  x\right)  -x\varkappa^{\prime\prime}\left(
x\right)  -\varkappa^{\prime}\left(  x\right)  =sy-\ln\varphi_{1}^{\prime
}\left[  \Theta\left(  sx\right)  \right]  .
\]
This equation for $\varkappa$ differs from Eq.~(\ref{A250}) only by the last
term on the right-hand side. So, as at $q_{m}=2$, we have asymptotically
$\psi_{0}\left(  s,x\right)  \sim x^{-1/2}T\left(  s,x\right)  \sim
x^{-1/2}\exp\left[  -\vartheta\left(  s,x\right)  \right]  $, where
$\vartheta\left(  s,x\right)  $ is given by Eq. (\ref{500}).

\section{Eigenfunction with minimum characteristic number at $s=0$}

\label{P}

At $s=0$, Eq. (\ref{140}) takes the form:%
\begin{equation}
T_{0}\left(  x\right)  =e^{-x}\left\{  1+\sqrt{x}\int_{0}^{\infty}\frac
{dy}{\sqrt{y}}I_{1}\left(  2\sqrt{xy}\right)  e^{-x}\varphi_{1}\left[
T_{0}\left(  y\right)  \right]  \right\}  . \label{P10}%
\end{equation}
Let us differentiate both the parts of this relation with respect to $x$. It is easy to check the identity:%
\[
\frac{\partial}{\partial x}\left[  e^{-x-y}\sqrt{\frac{x}{y}}I_{1}\left(
2\sqrt{xy}\right)  \right]  =-\frac{\partial}{\partial y}\left[  e^{-x-y}%
I_{0}\left(  2\sqrt{xy}\right)  \right]  .
\]
After differentiating and using this identity, we integrate by parts on the
right-hand side. The integrated term $e^{-x}$ on the lower limit of
integration, $y=0$, will be cancelled by the result of differentiating the
$e^{-x}$. Finally, we have
\begin{equation}
T_{0}^{\prime}\left(  x\right)  =e^{-x}\int_{0}^{\infty}dyI_{0}\left(
2\sqrt{xy}\right)  e^{-y}\varphi_{1}^{\prime}\left[  T_{0}\left(  y\right)
\right]  T_{0}^{\prime}\left(  y\right)  . \label{P20}%
\end{equation}
Comparing this with Eq. (\ref{2020}) at $s=0$, we see that $-T_{0}^{\prime
}\left(  x\right)  $ is the (unnormalized) eigenfunction of this equation
corresponding to the eigenvalue $\mu_{0}=1$. It is known from the theory of
linear integral equations \cite{zkkmrs68} that the eigenfunction corresponding
to the maximum characteristic number can be chosen to be real and positive
within the interval of integration. One can see that $-T_{0}^{\prime}\left(
x\right)  <0$ at any $x>0$, and the corresponding eigenvalue $\mu_{0}=1$ is
indeed a maximal one.


\begin{thebibliography}{99}
\bibitem {remark1}In fact, we applied a slightly different procedure. After
step (i) we knew only the asymptotic of $T\left(  s,x\right)  $ at
$\operatorname{Re}x\gg1$ and $\left\vert s-s_{c}\right\vert \ll1$. The
analytic continuation of $T\left(  s,x\right)  $ as a function of the variable
$s$ to the area $s<s_{c}$ has a singularity in terms of the variable $x$ at
large $x$. Therefore, to calculate $\rho\left(  \lambda\right)  $ at
$\lambda>\lambda_{c}$, one can use (\ref{340}). Then the asymptotics of
$\rho\left(  \lambda\right)  $ at $\lambda\rightarrow\lambda_{c}+0$ is defined
by the position of the singularity with respect to the variable $x$, and by
the behavior of $T\left(  -\lambda,x\right)  $ near this singularity.
\end{thebibliography}
\end{document}